\documentclass[twocolumn,superscriptaddress,english,prl,showpacs,longbibliography]{revtex4-2}
\usepackage[colorlinks=true,urlcolor=blue,citecolor=blue,linkcolor=blue]{hyperref} 
\usepackage{graphicx}
\usepackage{svg}
\usepackage{dcolumn}
\usepackage{bm}
\usepackage{color}
\usepackage{physics}
\usepackage{subfigure}
\usepackage{amsmath,amssymb,amsthm,mathrsfs,amsfonts,dsfont}
\usepackage{float}
\makeatletter
\let\newfloat\newfloat@ltx
\makeatother
\usepackage{algorithm} 
\usepackage{algorithmic}
\usepackage{pifont}
\renewcommand{\algorithmicrequire}{\textbf{Input:}}
\usepackage{balance}
\renewcommand{\algorithmicensure}{\textbf{Output:}}
\newcommand{\s}{\mathbf{s}}

\newcommand{\n}{\mathcal N}

\newcommand{\tp}{\widetilde {\mathcal P}}

\begin{document}

\title{Tensor Network Message Passing}

\author{Yijia Wang}
\affiliation{
 CAS Key Laboratory for Theoretical Physics, Institute of Theoretical Physics, Chinese Academy of Sciences, Beijing 100190, China
}
\affiliation{
School of Physical Sciences, University of Chinese Academy of Sciences, Beijing 100049, China}

\author{Yuwen Ebony Zhang}
\affiliation{Department of Physics and Astronomy, University College London,
Gower Street, London, WC1E 6BT, UK}

\author{Feng Pan}
\affiliation{
 CAS Key Laboratory for Theoretical Physics, Institute of Theoretical Physics, Chinese Academy of Sciences, Beijing 100190, China
}

\author{Pan Zhang}
\email{panzhang@itp.ac.cn}
\affiliation{
 CAS Key Laboratory for Theoretical Physics, Institute of Theoretical Physics, Chinese Academy of Sciences, Beijing 100190, China
 }
 \affiliation{School of Fundamental Physics and Mathematical Sciences,
 Hangzhou Institute for Advanced Study, UCAS, Hangzhou 310024, China}
 \affiliation{International Center for Theoretical Physics Asia-Pacific, Beijing/Hangzhou, China}

\begin{abstract}
When studying interacting systems, computing their statistical properties is a fundamental problem in various fields such as physics, applied mathematics, and machine learning. However, this task can be quite challenging due to the exponential growth of the state space as the system size increases. Many standard methods have significant weaknesses. For instance, message-passing algorithms can be inaccurate and even fail to converge due to short loops. At the same time, tensor network methods can have exponential computational complexity in large graphs due to long loops. This work proposes a new method called ``tensor network message passing.'' This approach allows us to compute local observables like marginal probabilities and correlations by combining the strengths of tensor networks in contracting small sub-graphs with many short loops and the strengths of message-passing methods in globally sparse graphs, thus addressing the crucial weaknesses of both approaches. Our algorithm is exact for systems that are globally tree-like and locally dense-connected when the dense local graphs have limited treewidth. We have conducted numerical experiments on synthetic and real-world graphs to compute magnetizations of Ising models and spin glasses, to demonstrate the superiority of our approach over standard belief propagation and the recently proposed loopy message-passing algorithm. In addition, we discuss the potential applications of our method in inference problems in networks, combinatorial optimization problems, and decoding problems in quantum error correction.
\end{abstract}
\maketitle

Consider \textit{statistical mechanics} problem defined on a graph $\mathcal G$ with $n$ vertices and a set of $m$ edges $\mathcal E$, and binary configurations $\s\in\{+1,-1\}^n$, which follow Boltzmann distribution
\begin{equation}
    P(\s)=\frac{1}{Z}e^{-\beta E(\s)},
\end{equation}
where $\beta$ is the inverse temperature, the energy function with external field $\{\theta_i\}$ is {$E(\s)=\sum_{(ij)\in\mathcal E} E_{ij}(s_i,s_j)+\sum_i\theta_i(s_i)$}, 
and $Z=\sum_{\s}e^{-\beta E(\s)}$ is the partition function.
Computing the macroscopic observables of the system such as the magnetizations and correlations are important problems in statistical physics, applied mathematics, and machine learning, and finding applications in 
inference and learning problems where the Boltzmann distribution naturally appears as the posterior distribution of Bayesian inference, 
in decoding error correction codes where signals can be reconstructed using marginals of the Boltzmann distribution, and in solving combinatorial optimization problems where the solutions map to typical samples of the Boltzmann distribution at zero temperature, and many others.

The computation of the local observables suffers from the large computational space which grows exponentially with the system size. The problem has the same complexity as computing the partition function and falls into the class of \#P problems in mathematics, so there is no polynomial algorithm that solves the problem exactly in general. Many methods have been proposed, including Markov Chain Monte Carlo (MCMC), message-passing algorithms, tensor networks, etc. 
MCMC~\cite{newman_monte_1999} is a general method for computing observables using unbiased samples, but the precision grows slowly with the number of samples hence is computationally expensive in large systems. Moreover, for systems with a complicated landscape, MCMC has the autocorrelation issue.

Message passing algorithms such as belief propagation~\cite{yedidia_understanding_2001} and survey propagation~\cite{mezard_analytic_2002} are celebrated methods in statistical physics, and also play an important role in decoding low-density parity-check code~\cite{gallager1962low}, solving random combinatorial optimization problems~\cite{mezard_analytic_2002}, detecting structures and signals in large networks~\cite{decelle_asymptotic_2011}, etc. It is closely related to the Bethe mean-field approximations~\cite{bethe_statistical_1935}, also known as the cavity method in statistical physics~\cite{mezard_bethe_2001,mezard_cavity_2003}.
Message-passing algorithms usually have low computational costs, but the performance heavily relies on the topology of the system, {working} well only on locally tree-like graphs without many short loops. 
Many efforts have been devoted to extending the message-passing algorithm to systems with short loops~\cite{kikuchi_theory_1951,yedidia_generalized_2000,chertkov_loop_2006,montanari_how_2005,pelizzola_cluster_2005,parisi_loop_2006,zhou_region_2012,cantwell2019message,Kirkley_2021}, however, so far the extensions have very limited success, only dealing with very short loops inside a small region of the graph.
Tensor network methods~\cite{schollwock_density-matrix_2011,orus_practical_2014} are powerful on graphs full of short loops, particularly on lattices, especially in two dimensions, because tensor contractions can eliminate short loops with various sizes efficiently~\cite{pan_contracting_2020}. However, for systems without translational invariance tensor networks only apply to small systems, due to the fast growth of the computational complexity with system size when there are long loops.
\begin{figure*}[t]
\centering
    \centering
    \includegraphics[width=0.98\textwidth]{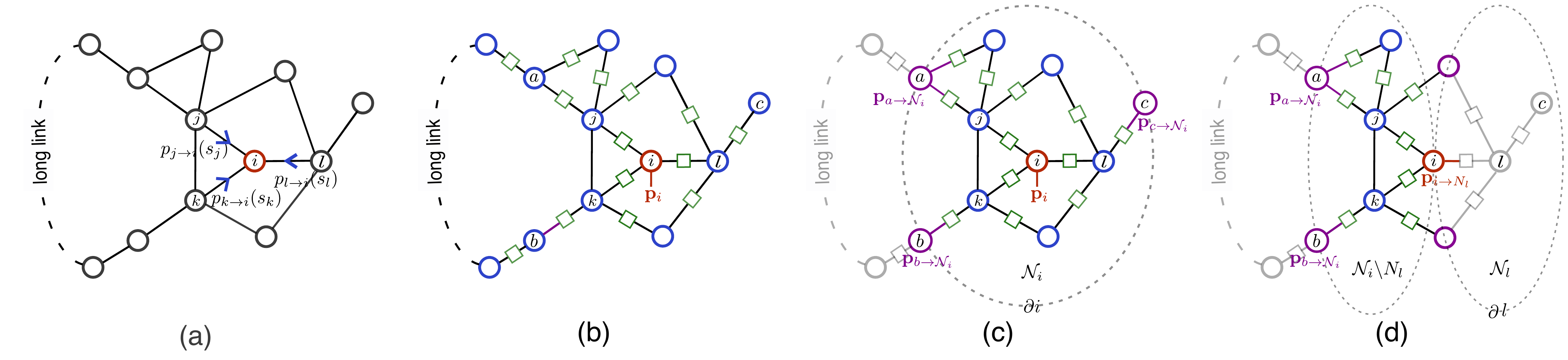}
\caption{Pictorial illustration of belief propagation (BP) (a), tensor network contraction (b), and tensor network message passing (TNMP) (c) for computing the marginal probability of node $i$ in an Ising model on a graph. In (b), (c), and (d), circles and squares are tensors converted from the Ising model. The shaded area in each figure denotes the part of the graph that is involved in computing the marginal of $i$. In BP, it is computed using information from its direct neighbors; in exact tensor network contractions, it includes all tensors in the network; and in TNMP, it evolves a pre-defined neighborhood $\mathcal N_i$. In panel (d) The update of $p_{i\to\n_l}$ in TNMP by contracting tensors inside $\n_i\backslash \mathcal N_l$ and the cavity tensors (the purple ones).}
\label{fig:diagram}
\end{figure*}

In this work we propose the tensor network message passing (TNMP) method to combine the advantage of tensor networks in contracting short loops and the advantage of message passing in iterating over long loops, addressing the issues for both of them. Our methods rely on the arbitrary tensor network contraction methods developed recently in the context of classical simulation of quantum computers~\cite{pan_simulation_2022,pan_solving_2022,pan_contracting_2020,xu2023herculean}, with computational complexity depending on the treewidth of the neighborhood graph rather than the number of nodes. Thus TNMP can work with a neighborhood that are much larger than existing loop message-passing methods. Using Ising and spin glass models on synthetic and real-world networks, we demonstrate the superiority of our method to belief propagation, MCMC, and the recently proposed loopy message passing algorithm. In what follows, using a concrete example illustrated in Fig.~\ref{fig:diagram},
we first review message passing and tensor network, then introduce our method.

\paragraph {Message passing algorithm---} 
As shown in Fig~\ref{fig:diagram}(a), in belief propagation the marginal distribution $p_i(s_i)=\sum_{\s\backslash s_i}P(\s)$ is computed using information passed from neighbors, involving a small neighborhood subgraph composed of $i$ and its direct neighbors. 
Without loss of generality let us consider the Ising model, with $E_{ij}(s_i,s_J)=-J_{ij}s_i s_j$ and $\theta_i(s_i) = -h_is_i$, where $h_i$ is the external field. Then the detailed computation can be written as 
$$ p_{i}(s_i) = \frac{e^{\beta h_is_i}}{Z_i}\prod_{k\in\partial i}\sum_{s_k}e^{\beta J_{ik}s_is_k}p_{k\to i}(s_k).$$
Where $p_{k\to i}(s_k)$ is the cavity message indicating the marginal probability of node $k$ taking value $s_k$ when node $i$ is removed from the graph and can be determined using cavity messages sent from the neighbors of $k$, but without $i$, in the same manner of computing the marginal. 
The key assumption of BP is the conditional independence, i.e. $p_{j\to i}(s_j)$, $p_{k\to i}(s_k)$, and $p_{l\to i}(s_l)$ are independent. This assumption is correct only when the neighbors are not connected to each other via other nodes in the graph. 
However, when the graph contains many short loops, belief propagation has a long-standing problem of suffering from poor accuracy, because when the neighbors are connected by short loops (e.g. as shown Fig.\ref{fig:diagram}(a)), the conditional independence apparently does not hold, resulting in an inaccurate marginal computation.

\paragraph{Tensor networks---} 
On the opposite, tensor networks are particularly good at eliminating short loops using tensor contractions. It maps the computation of the partition function $Z=\sum_{\s}e^{-\beta E(\s)}$ to the contraction of a tensor network with the same shape (see e.g. ~\cite{pan_contracting_2020,PhysRevB.104.075154}).
Local observables such as the marginal distribution of a node $i$, can be computed in the same way, e.g. $p_i(s_i)=Z(s_i)/\sum_{s=\pm 1}Z(s)$, where $Z(s_i)$ is the partition function given a configuration of node $i$ as $s_i$. The picture is depicted in Fig.~\ref{fig:diagram}(b), where the circuits represent diagonal tensors with only two non-zero elements, as 
$\raisebox{-3.8ex}{\includegraphics[scale=0.4, trim={0cm 0.0cm 0cm 0cm}, clip]{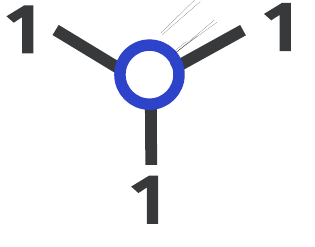}}= e^{\beta h_i}$,  $\raisebox{-3.8ex}{\includegraphics[scale=0.4, trim={0cm 0.0cm 0cm 0cm}, clip]{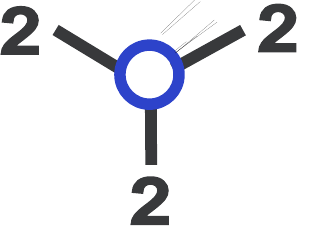}}=e^{-\beta h_i}$, and $0$ for all other tensor elements, and squares are $2\times 2$ matrices encoding energy terms as $
\raisebox{-1.5ex}{\includegraphics[scale=1, trim={0cm 0.0cm 0cm 0cm}, clip]{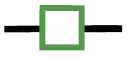}}= \left(\begin{array}{cc} e^{\beta J_{ij}} & e^{-\beta J_{ij}} \\  e^{-\beta J_{ij}} & e^{\beta J_{ij}} \end{array}\right)$ .
    It is clear from Fig.~\ref{fig:diagram}(b) that the computation of one node's marginal involves all the nodes in the graph. This arises an issue that it only works for small systems in general, as the exact contraction of the whole tensor network has computational complexity exponential in the treewidth of the graph~\cite{markov2008simulating}. For large graphs, one can not even store the intermediate tensor during the contraction. 
To illustrate this limitation, consider a concrete example on an infinite Caley tree (Bethe lattice) where BP is asymptotically exact. However, due to the long loops that can not be contracted immediately, contraction of any two tensors would result in a tensor with a larger dimension and the space complexity of exact tensor network contraction will inevitably grow to infinity.
Another limitation of tensor networks in computing the local observables is that direct contracting the overall tensor network only gives one local observable (e.g. magnetization of one node), so one has to repeat the whole-tensor-network contraction for each node.

\paragraph{Tensor Network Message Passing (TNMP) ---} In this work we propose to combine message passing and tensor networks in such a way that tensor network contractions are responsible for contracting short loops while message passing is responsible for long loops. We use an example to introduce our method. 

The marginal computation of the proposed method is depicted in Fig.~\ref{fig:diagram}(c), a shaded area $\mathcal N_i$, which we term as \textit{neighborhood} of $i$, is involved in computing marginal probability $p_i(s_i)$. 
The tensors inside $\n_i$ such as $j,k$ and $l$, and cavity tensors on the boundary such as $a$, $b$, and $c$ are contracted in computing $p_i$. While tensors (the gray ones in the figure) outside the $\n_i$ do not contribute to the computation of $p_i$. 
The tensors on the boundary provide an environment to the contraction of $\n_i$, i.e. contraction results of all the tensors of the tensor networks excluding those in $\n_i$. However, computing the exact environment is equivalent to contracting exactly the whole tensor network. In this work, we introduce a rank-one approximation of the environment as a tensor product of all \textit{cavity tensors} $p_{a\to\n_i}$, $p_{b\to\n_i}$, and $p_{c\to\n_i}$, which assumes independence between the cavity tensors.
The approximation is exact if $a$, $b$, and $c$ are not connected by tensors outside $\n_i$, and is a good approximation if they are connected only by long loops via tensors outside $\n_i$. We compute the cavity tensors iteratively, analogous to computing cavity messages in message-passing algorithms. An example is given in Fig.~\ref{fig:diagram}(d), $p_{i\to\n_l}$ is computed using the tensor contraction of tensors in $\n_i$ excluding tensors inside $\n_l$, and the tensors on the boundary become cavity tensors.

To determine the neighborhood, we propose to generate $\n_i$ by progressively including neighboring tensors, subject to that the minimum distance $\min_{(ab)}d_{ab}(\partial \n_i)$ between all pairs of tensors $(a,b)$ on the boundary $\partial \n_i$ not smaller than a given value $R$. Here the distance $d_{ab}(\partial \n_i)$ between tensors $a$ and $b$ is defined as the length of the shortest path connecting $a$ and $b$ only via the tensors outside $\n_i$ (e.g. via the gray part of the tensor network in Fig.~\ref{fig:diagram}(c) and (d)). 
The idea behind the neighborhood generation is intuitive: the longer $d_{ab}$, the weaker cavity tensors $a$ and $b$ are correlated, when $\n_i$ is removed from the graph. 
In a special case when the tensor network is a tree, for any choice of $\n_i$, the tensors on the boundary are not connected without passing node $i$, i.e. with $d_{ab}(\partial \n_i)=\infty$, so we can generate $\n_i$ simply using the direct neighbors of $i$, and TNMP reduces to belief propagation in this case. 
For a general graph with many short loops and long loops as depicted in Fig.~\ref{fig:diagram}(a), a larger minimum distance results in a more accurate computation of local variables, because the cavity tensors on the boundary have weaker correlations, and $\n_i$ we generate is larger and the tensor contraction involves more tensors. If $\n_i$ span all the tensors in the network (i.e. $|\n_i|=n$), then $d(\partial \n_i)=\infty$ and the marginal computation is exact, which is actually the conventional tensor network method. 
In this sense, our algorithm generalizes both the tensor network method and belief propagation.

The computational complexity of computing the local variable by contracting the neighborhood depends on the treewidth of the neighborhood graph. If treewidth is small, we use exact tensor network contraction by finding a good contraction order~\cite{kourtis_fast_2019,gray_hyper-optimized_2021,pan_simulation_2022,pan_solving_2022,kalachev2021multi}. If the treewidth is large, we employ an approximate contraction algorithm that works for arbitrary tensor networks~\cite{pan_contracting_2020}. We note that in computing local expectations e.g. marginals (or magnetizations) for all variables, TNMP only needs to converge once, giving all consistent environments which can be further used to compute local observables by contracting local tensor networks. 

\begin{figure}
\centering
\includegraphics[width=0.98\columnwidth]{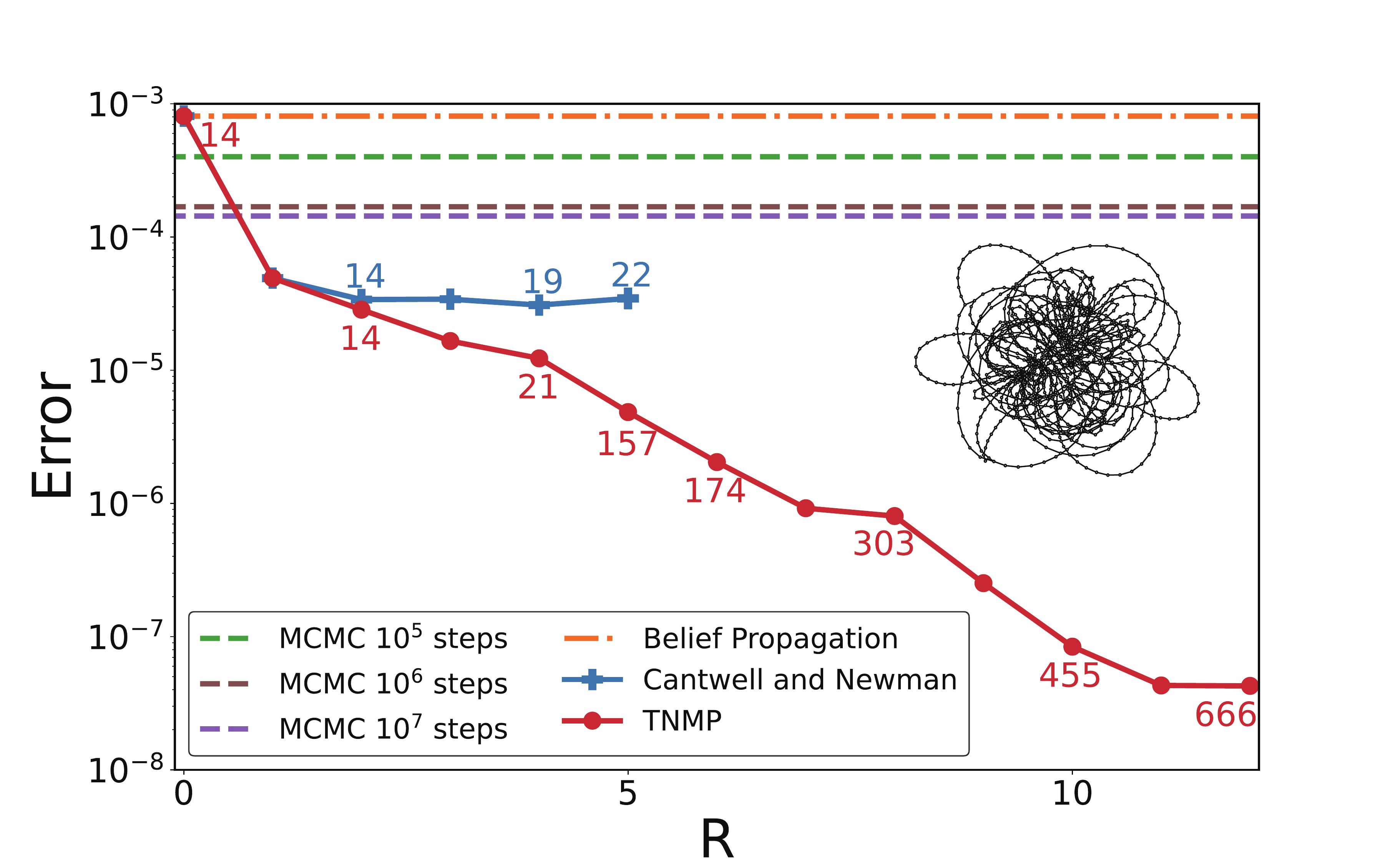}
\includegraphics[width=0.98\columnwidth]{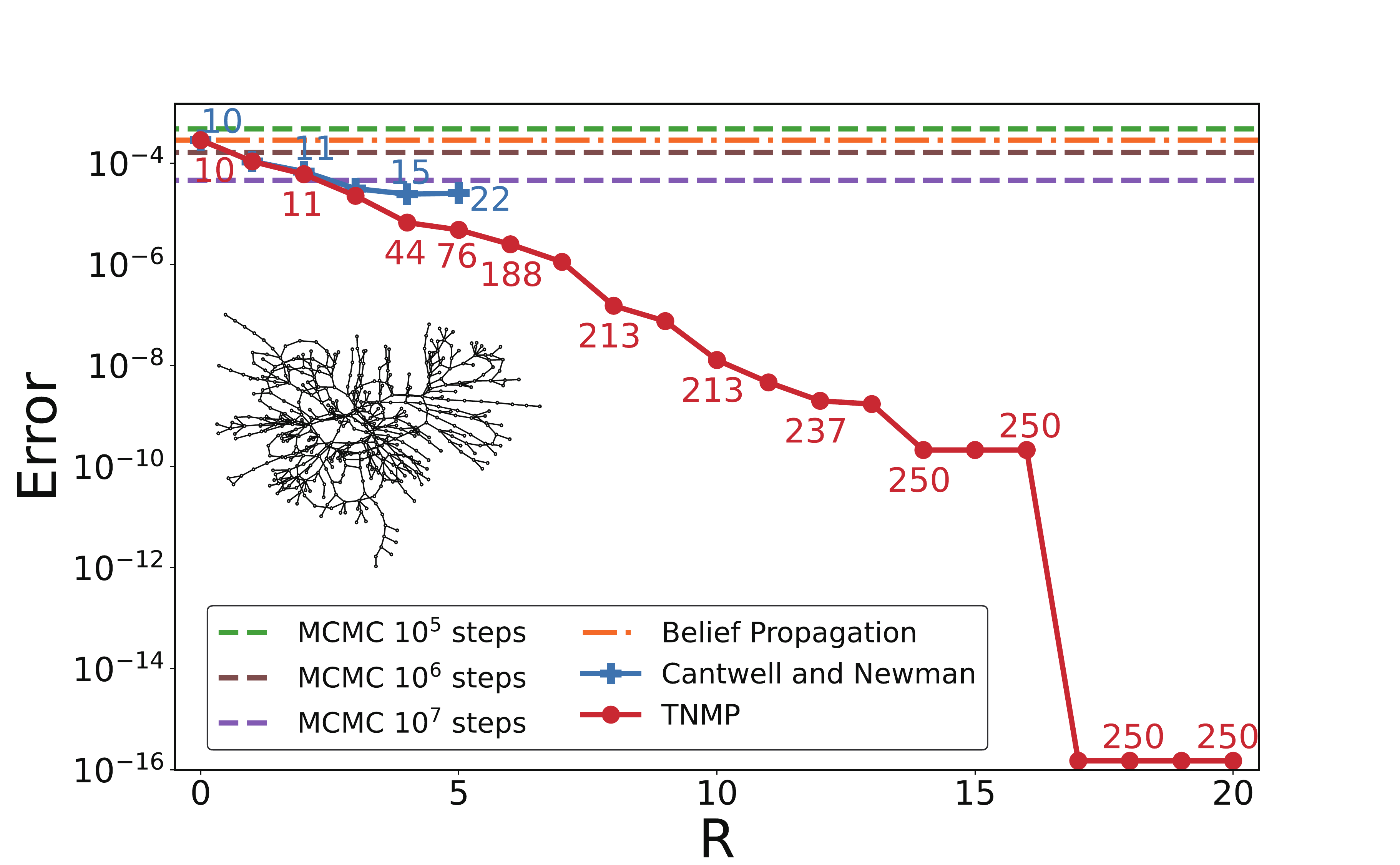}
\caption{\label{fig:eva}
({\textit Top}): Error of magnetizations $E=\frac{1}{n}\sqrt{\sum_{i=1}^n{(M_i - M^{\mathrm{exact}}_i)^2}},$ given by various methods for a ferromagnetic Ising model on a synthetic graph (illustrated in the inset) generated using the model of~\cite{williamson2020random}. The graph has $n=1000$ nodes and contains a random graph and cliques on top of it. The random graph has an average degree $d=2$, the cliques have a size ranging from $2$ to $9$, and the number of them is $[36,12,6,3,2,1,1,1]$ respectively.
The temperature of the model is fixed at $T=2.0$ and the external field $h=0.1$. $R$ in the X-axis is the minimum distance between all pairs of nodes on the boundary of the neighborhood in our method TNMP (see text) and denotes the maximum length of the path under consideration between the neighbors of a node in Cantwell and Newman's method~\cite{cantwell2019message,Kirkley_2021}. 
The numbers labeled in the figure indicate maximum neighbor size $\max_{i}|\n_i|$ with a given $R$.
({\it Bottom}): Errors of magnetization obtained by various methods for a random-interaction and random-field spin glass model on the real-world network of bus power system~\cite{10.1145/2049662.2049663} at a fixed temperature $T=2.0$. The coupling $J$ are chosen as $\pm 1$ with equal probability and the external field $h_i$ for node $i$ follows $h_i \sim  N(0,\frac{\pi}{200})$.
}
\end{figure}

\paragraph{Numerical experiments}
We evaluate our method using two examples. The first one is an Ising model on a synthetic graph containing both random links and cliques as proposed in \cite{williamson2020random}. The generation process creates a random graph with average degree $d$, then draws cliques with different sizes following a given distribution. The graph generated in this way is sparse globally and dense locally, hence is challenging to both the canonical tensor network method and belief propagation algorithm. 
In our evaluation, we randomly generated a network with $n=1000$ nodes, random links with $d=2$, and cliques with sizes ranging from $2$ to $9$. A scatch of the graph is shown in the inset of Fig.~\ref{fig:eva} top. The graph is generated in such a way that it is still possible to compute the exact magnetizations $M^{\mathrm{exact}}$ using the contraction of the whole tensor network, with the help of the dynamic slicing and multi-GPU computation which are modern techniques developed very recently in the large-scale quantum circuit simulations~\cite{pan_simulation_2022,pan_solving_2022}.
In Fig.~\ref{fig:eva} top, based on the exact results, we show the error of each node's magnetization given by our method, compared with the error given by the recent loopy message passing method of Cantwell and Newman~\cite{cantwell2019message}, and MCMC. 
In the figure, $R$ in the X-axis controls the size of the neighborhood involved in the magnetization computation. In our method, $R$ is the minimum distance $\min_{(ab)}d_{ab}(\partial \n_i)$ between all pairs of tensors on the boundary of the neighborhood. 
In Cantwell and Newman's method,  $R$ is the maximum length of the path under consideration between the neighbors of a node. 

From the figure, we can see that the error of our method decreases monotonically with $R$. With $R=0$, our method reduces to belief propagation and gives the same result. With $R=2$, although the maximum neighborhood size of our method and~\cite{cantwell2019message} are the same, our method gives better results. The difference becomes larger with $R=3$,$4$, and $5$. 
We note that the computational cost of Cantwell and Newman's~\cite{cantwell2019message} method (without employing Monte-Carlo sampling) is exponential in the neighborhood size (as labeled in the figure), which increases rapidly with $R$ so in the evaluation, we restrict it to $5$.
For our method, the computational cost is related to the treewidth of the neighborhood, rather than directly relating to its size, hence works for a large $R$.
For contracting the neighborhood sub-tensor-networks, we use exact tensor network contraction with $R< 9$. For $R\geq 9$ the neighborhood sub-tensor-networks are very large so we use CATN method~\cite{pan_contracting_2020} to contact them. We can see that our method works to the maximum neighborhood size as large as $666$, giving an error smaller than $10^{-7}$.
We also include the MCMC results with different numbers of update steps. In each step, $n$ spins are randomly chosen one by one and updated sequentially according to the Metropolis-Hasting algorithm. We see that the error decreases slowly with a larger number of steps, obtaining $10^{-4}$ with $10^7$ steps.

In Fig.~\ref{fig:eva} bottom we evaluate our method using a spin glass model (with random $\pm 1$ couplings and random fields) on the real-world electric power grid network~\cite{10.1145/2049662.2049663} containing $n=494$ nodes.
For this graph, we obtain exact magnetizations and evaluate the error of magnetization. With all $R$ values, in TNMP we can always contract sub-tensor networks exactly, even when the largest neighborhood contains $250$ nodes. 
In the figure, we can see that TNMP gives much smaller errors than belief propagation, the method in~\cite{cantwell2019message}, and MCMC with a large number of updating steps.
Moreover, when $R$ is larger than $16$, the error of our method drops to the rounding error, indicating that all the loops have been included in the neighborhood, and TNMP is exact even when the maximum neighborhood size is smaller than the number of nodes $494$.

\paragraph {Discussions---}\label{sec:dis}
We have introduced the TNMP method that combines tensor networks and message passing. Although we demonstrate our method using models of statistical mechanics, our methods immediately find potential applications in a broad area in different fields of science. 
It is straightforward to apply our method to computing the spectra for sparse matrices and the percolation problems~\cite{cantwell2019message}, it also applies to the inference problems and community detection~\cite{decelle_inference_2011,zhang_scalable_2014} in real-world networks, particularly suitable for networks which are globally sparse and with local motifs. 
Since our method does not restrict the local tensor contractions for real numbers, we can easily extend TNMP to the complex field for computing expectations in quantum systems~\cite{sahu2022efficient}, or even extend using other kinds of algebra e.g. the Tropical algebra in the semi-ring~\cite{liu_tropical_2021}. 
We can also extend survey propagation (SP)~\cite{mezard_analytic_2002,braunstein_survey_2005} with tensor networks in solving constraint satisfaction problems. Similar to BP, we can map SP to a local tensor network contraction and extend the local contraction to a large neighborhood. This approach would leverage SP and one-step replica-symmetry breaking methods to constraint satisfiability problems on loopy and real-world large graphs.

Another interesting application is decoding of quantum error correction code~\cite{nielsen2002quantum}. Due to the degeneracies, decoding requires summing over all elements of the stabilizer group. And due to the commutation relation of stabilizers, the factor graph of the quantum code naturally contains many short loops, in contrast with classical error correction code (such as the low-density parity-check code) which can be designed to almost contains no short loops. BP decoder and its variants so far do not perform well in quantum error correction code. We expect that our TNMP approach can alleviate the problem of short loops for message-passing decoders in surface code and quantum low-density parity check code.

\begin{acknowledgments}
A \textsc{python} implementation and a Jupyter notebook tutorial of our algorithm are available at \cite{tnmp}
    We thank Federico Ricci Tersenghi, Chuang Wang, and Haijun Zhou for helpful discussions on the manuscript, and Alec Kirkley for discussing and for sharing the code and data of Ref.~\cite{Kirkley_2021}. P.Z. acknowledges the WIUCASICTP2022 grant and Project 11747601 and 11975294 of NSFC.
\end{acknowledgments}

\bibliography{tnmp.bib}
\onecolumngrid
\renewcommand{\theequation}{S\arabic{equation}}
\setcounter{equation}{0}
\renewcommand{\thefigure}{S\arabic{figure}}
\setcounter{figure}{0}

\balance
\section{Computing marginal probabilities with tensor network contractions}
Without loss of generality, let us consider the Ising model with binary variables defined on a graph $G=(\mathcal{V,E})$, with configurations following Boltzmann distribution
\begin{equation}
    P(\s)=\frac{1}{Z}e^{-\beta[\sum_{(ij)\in\mathcal{E}} E_{ij}(s_i,s_j)+\sum_{i \in \mathcal{V}}\theta_i(s_i)]},
\end{equation}
where $E_{ij}(s_i,s_J)=-J_{ij}s_i s_j$ and $J_{ij}$ are coupling constants, $\theta_i(s_i) = -h_is_i$ with $h_i$ denoting the external fields and $Z=\sum_{\s}e^{-\beta E(\s)}$ is the partition function. For any subset of vertices $\mathcal{V}_i = \left\{ i_1,i_2,\cdots ,i_m\right\} \subset \left\{ 1,2,\cdots, n\right\}$, the marginal probability $\s_{\mathcal{V}_i}$ is defined as
\begin{equation}
    \begin{split}
    \label{eq:marginal}
    P(\s_{\mathcal{V}_i} = x_{\mathcal{V}_i})
    =&\frac{1}{Z} \sum_{\s \backslash\s_{\mathcal{V}_i} } e^{-\beta[\sum_{(ij)\in\mathcal{E}} E_{ij}(s_i,s_j)+\sum_{i \in \mathcal{V}}\theta_i(s_i)]}\\
    =&\frac{1}{Z} \sum_{\s \backslash\s_{\mathcal{V}_i} }\prod_{(ij)\in\mathcal{E}}e^{\beta J_{ij}s_is_j} \prod_{i \in \mathcal{V}}e^{\beta h_is_i}
    \end{split}
\end{equation}

The energy function of the Boltzmann distribution is always easy to compute, i.e. with polynomial time. But the properties of the Boltzmann distribution, e.g. observables like marginal distributions, correlations, and expected energy are hard to obtain because the explicit probability values of each configuration $\s$ are not known due to the hardness of computing the partition function, i.e. the normalization of the Boltzmann distribution, $Z$. It is well known that computing $Z$ is a \#P problem so there is no polynomial algorithm to solve it. A na\"ive way  for computing $Z$ is by enumerating all possible configurations but it only works for a system with a number of variables smaller than $20$ or so. The tensor network avoided enumerating the configuration of variables. Instead, it exploits the connectivity graph's topology and structure and usually reduces the computational cost heavily. As a simple example, consider computing a partition function of an Ising model with $30$ spins on a chain. It is well known that the exact computation of the partition function of the model is easy e.g. by ``eliminating variables'' from one end of the chain to the other end, hence is not necessary to enumerate all possible configurations. Actually, this variable elimination strategy can be considered as a special example of a tensor network contraction algorithm for computing partition function for a statistical mechanics model with discrete variables.

In principle, any distribution over discrete variables can be considered as a tensor in a finite dimension. For example, $P(\s)$ is a ``probability vector'' $\mathcal P$ with dimension $2^n$ and can be reshaped into a $n$-way ``probability tensor'' with dimension $\underbrace{2\times 2\times 2\times\cdots\times 2}_{n}$. The unnormalized version $e^{-\beta[\sum_{(ij)\in\mathcal{E}} E_{ij}(s_i,s_j)+\sum_{i \in \mathcal{V}}\theta_i(s_i)]}$ can be treated in the same way as a $n$-way tensor $\tp$. The difference between $\mathcal P$ and $\tp$ is the sum of all elements, i.e. $\ell_1$ of the tensor: $|\mathcal P|_1=1$ while $|\tp|_1=Z$. This indicates that the partition function can be computed using the inner product between $\tp$ and an all one vector 
$\{\underbrace{1,1,1,\cdots,1,1,1}_{2^n}\}$. This computation seems impossible but sometimes can be computed efficiently making use of the fact that $\tp$ can be formulated as a tensor network composed of diagonal ``field tensors'' corresponding to the variables and \textit{Boltzmann matrices} corresponding to the interaction on each edge.
The Boltzmann matrix on an edge $(ij)$ with interaction $J_{ij}$ is constructed as
\begin{equation}
    B^{ij} = 
    \begin{pmatrix}
        e^{\beta J_{ij}} & e^{-\beta J_{ij}} \\
        e^{-\beta J_{ij}} & e^{\beta J_{ij}},
    \end{pmatrix}
\end{equation}
and the field tensors for each variable with the external field $h_i$ is written as 
\begin{equation}
    F^i_{i_1,i_2,\cdots i_d} = 
    \begin{cases}
        e^{\beta h_i}& i_1=i_2=\cdots=i_d=1\\
        e^{-\beta h_i}& i_1=i_2=\cdots=i_d=2\\
        0& \text{otherwise}
    \end{cases}
\end{equation}
where $d$ is the degree of the vertex $i$.

\begin{figure}
\centering
\includegraphics[width=0.98\columnwidth,trim=20 10 10 5,clip]{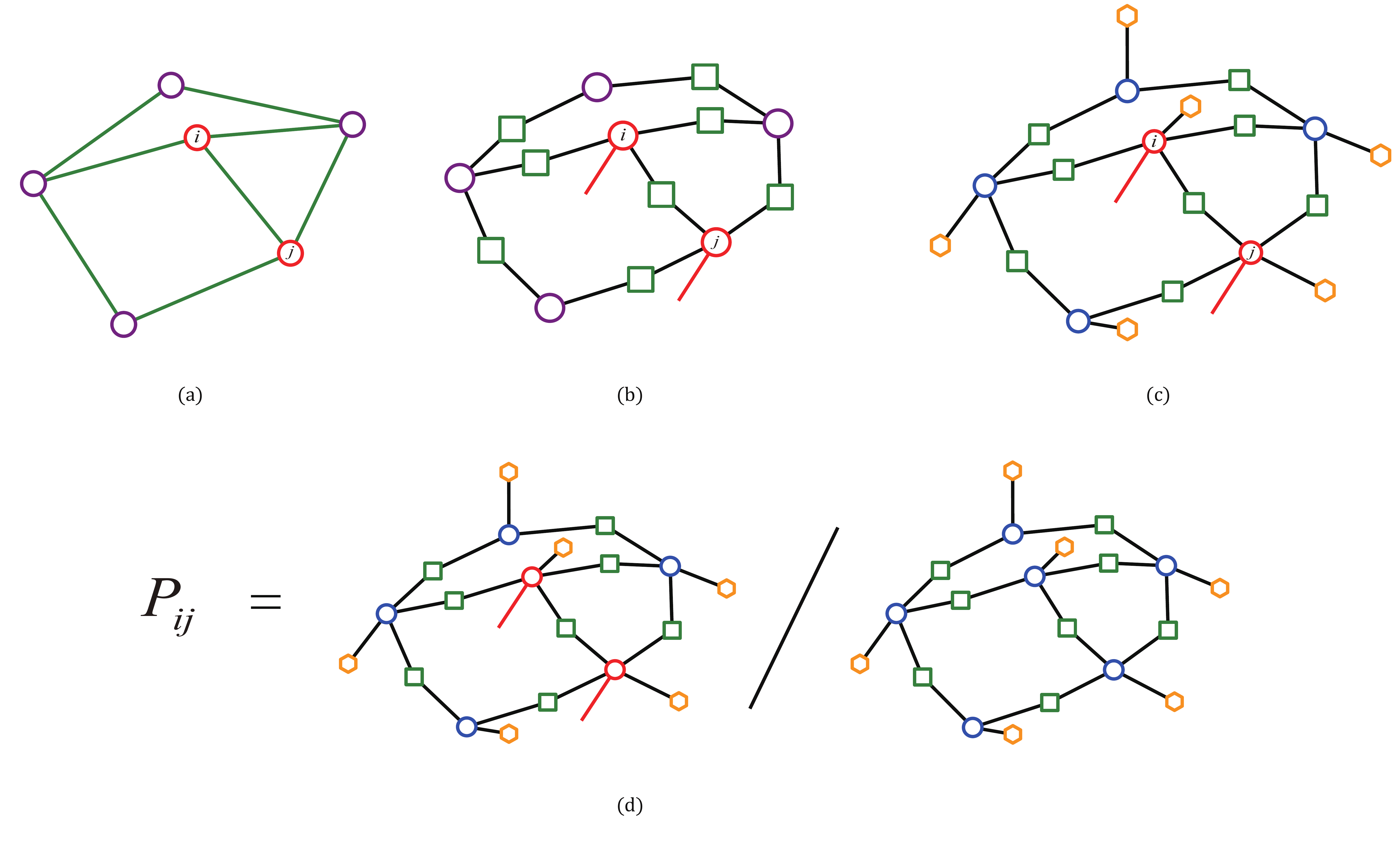}

\caption{Mapping the marginal probabilities in the Ising model to tensor networks. (a) The original graph $G$ upon which the model is defined. (b) The tensor network correspondence of the un-normalized numerator $Z(s_i,s_j)$, where the purple circles represent $F^i$, the green squares are $B^{ij}$ and the two red legs on $i,j$ are the open indicators of the result tensor. (c) Another form of $T$, where every $F^i$ in (b) is written as a copy tensor (represented by a blue circle) with a vector $f^i$  (represented by an orange hexagon) on it. (d) The tensor network representation of $P(s_i,s_j)$, which is the quotient of the tensor networks of $Z(s_i,s_j)$ and $Z$.
}
\label{fig:GTT}
\end{figure}

Under this construction, the partition function can be computed by contracting a tensor network with exactly the underlying connectivity graph of the Ising model, with variables replaced by the field tensor and edges replaced by the Botzmann matrix. For the same construction, the normalization $Z(x_{\mathcal{V}_i})$ of the marginals in Eq. \ref{eq:marginal} can be computed using the contraction  of a tensor network $T$ composed of $B^{ij}$ and $F^i$. As shown as an example in Fig. \ref{fig:GTT}(b), where $T$ has the same topology with graph $G$ - the edge $(i,j)$ in $G$ corresponds to $B^{ij}$ in $T$ and the vertex $i$ corresponds to $F^i$ which can be further written as a copy tensor with a leg shared with a vector $f^i=(e^{\beta h_i},e^{-\beta h_i})$ in Fig. \ref{fig:GTT}(c).
In particular, when $\mathcal{V}_i$ is an empty set, $Z(x_{\mathcal{V}_i})$ is reduced to the partition function $Z$, we can get the normalized marginals by quotient the two tensor networks shown in Fig. \ref{fig:GTT}(d).

\subsection{The TNMP equations}
\begin{figure}
\centering
\includegraphics[width=0.98\columnwidth,trim=0 10 0 20,clip]{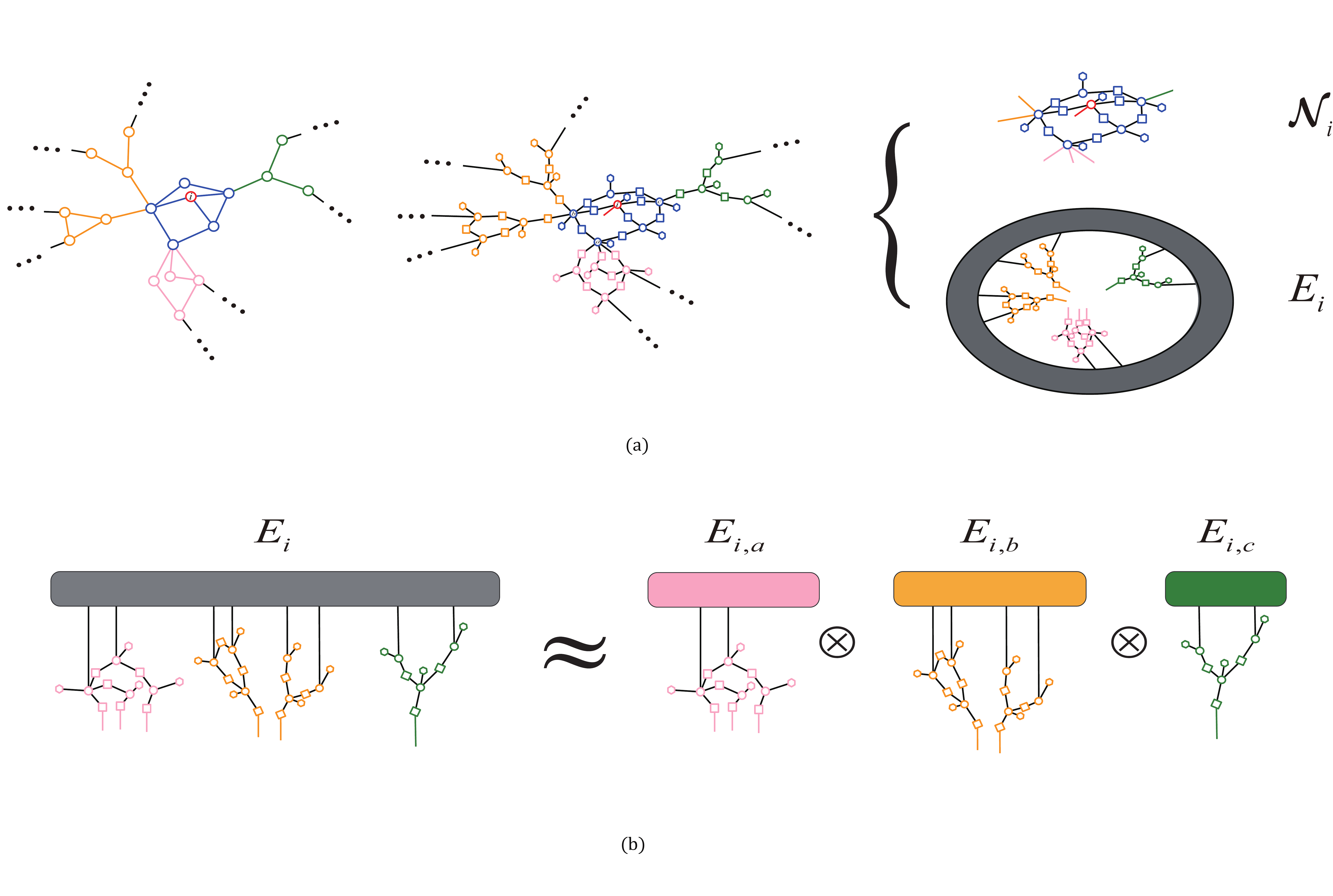}

\caption{Neighborhood-environment division and the 
 tensor product approximation of the environment. (a) The local sub-graph of the red vertex $i$, the corresponding local tensor network for $P_i(s_i)$ and its decomposition, where the grey circle in $E_i$ represents the tensor obtained by contracting the whole remaining tensor network outside our focus. (b) The 
 direct product approximation of the environment, where the $E_{i,a}$, $E_{i,b}$ and $E_{i,c}$ represent the environment tensors contracted from the three disconnected sub-networks which make up the whole environment under hypothetical ideal conditions.
}
\label{fig:NEP}
\end{figure}

To obtain exactly the single-node marginal $P_i(s_i)$ for node $i$, we need to contract the whole tensor network $T$. If we need only an approximate estimate of the marginal, it is usually not necessary to contract all the tensors in the tensor network which is computationally heavy. Observe that the tensors that are closer to the node $i$ usually contribute more significantly to the marginal. (As illustrated in Fig. \ref{fig:NEP}(a), the tensor network $T$ is divided into two parts - the \textit{neighborhood} of $i$, denoted as $\mathcal N_i$, and its complement, \textit{the environment}, denoted by $E_i$.)
So a na\"ive way of computing the marginal is that we can ignore the tensors far away from the node $i$ and only consider the contraction over the tensors around $i$, i.e. neighborhood of $i$. This approach converts the global computation of the marginal to a local computation and thus heavily reduces the computational complexity. However, since it completely ignores the tensors in $E_i$, the computation is not accurate, unless the size of $\mathcal N_i$ is large enough and $E_i$ becomes a non-sensitive boundary condition to $\mathcal N_i$. A more clever approach is efficiently approximating $E_i$ rather than completely ignoring it. 
As a generalization of the approximation in message passing algorithms, we assume that $E_i$ is composed of several disconnected sub-networks, each of which is connected to only one tensor on the boundary of $\mathcal N_i$, as shown in Fig. \ref{fig:NEP}(b). In other words, the environment tensor is assumed to factorizing into tensor product of several smaller tensors. If the actual network conforms to this assumption, (e.g. the green, yellow, and pink rectangles represent independent environment tensors from three different ``directions" as shown in the figure), the factorization of the environment tensor is exact.
Otherwise, it is a tensor product approximation of the environment tensor, e.g. the gray rectangle in the figure as $$E_i\approx E_{i,a}\otimes E_{i,b}\otimes E_{i,c}.$$
Apparently, the fewer (and weaker) connections between the  sub-networks, and the longer the connections are, the more accurate the tensor product approximation will be.

\begin{figure}
\centering
\includegraphics[width=0.98\columnwidth,trim=50 20 0 10,clip]{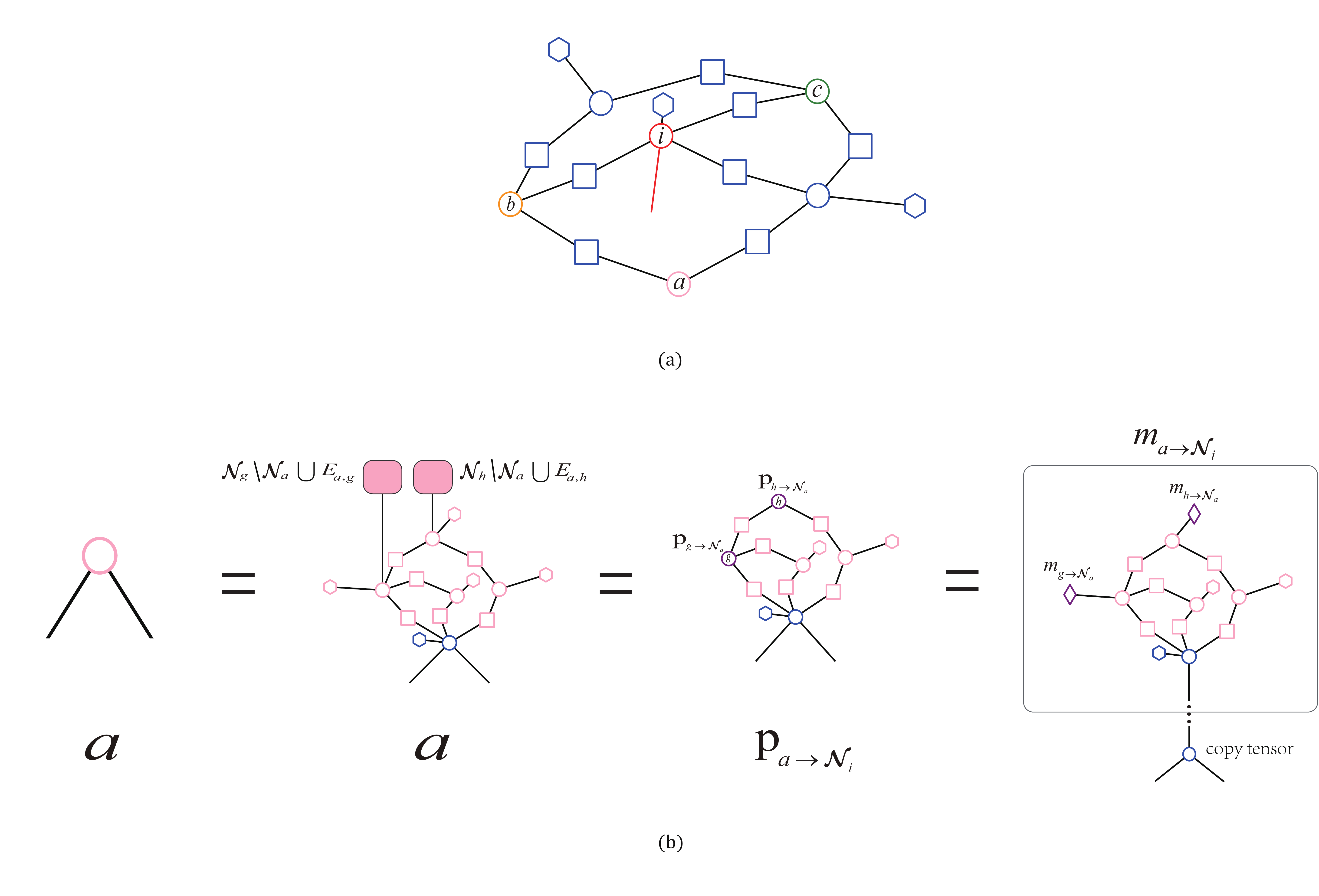}

\caption{Cavity approximation in the marginal calculation. (a) The local tensor network to be contracted in $P_i$ calculation, where the tensors $a$, $b$ and $c$ are the contraction of the environment sub-network in Fig. \ref{fig:NEP}(b) and the boundary tensor at the original position. (b) The cavity approximation of $a$, where the pink part of the network represents the cavity sub-network $\mathcal{N}_a \backslash \mathcal{N}_i$, the blue part represents the copy tensor and the field vector $f^a$ in $\mathcal{N}_i$, $p_{a \rightarrow \mathcal{N}_i}$, $p_{\text{g} \rightarrow \mathcal{N}_a}$ and $p_{h \rightarrow \mathcal{N}_a}$ represent the cavity tensors and $m_{a \rightarrow \mathcal{N}_i}$ represents the message vector.
}
\label{fig:Cavity}
\end{figure}

With the tensor product approximation, we can contract the environment into the boundary tensors of $\mathcal{N}_i$, then the calculation of $P_i(s_i)$ of Fig. \ref{fig:NEP} becomes the contraction of the local tensor network  $\mathcal{N}_i$ shown in Fig. \ref{fig:Cavity}(a) with all the environment information contained in the boundary tensors $a$, $b$ and $c$.

Although the entire tensor network is separated into local sub-networks, it is still computationally heavy to contract sub-networks e.g. $a$, $b$, and $c$ in the figure. In the spirit of the message-passing algorithms, we compute them iteratively as a function of cavity tensors which are in turn determined by an iterating algorithm. Similar to the computation of marginals, We first assume that the remaining network after removing $\mathcal{N}_a$ from $E_{i,a}$, represented as the pink rectangle in Fig. \ref{fig:NEP}(b) and Fig. \ref{fig:Cavity}(b), can also be decomposed into several disconnected sub-sub-networks, each of which is then contained in one of the boundary tensors of $\mathcal{N}_a$ like $\text{g}$  and $h$ in Fig. \ref{fig:Cavity}(b). As shown in Fig. \ref{fig:Cavity}(b), we make the same approximation on $\text{g}$ and $h$, then $E_{i,a}$ is approximated as the cavity tensor $p_{a\to\n_i}$, which is contracted from the local tensor network $\mathcal{N}_a \backslash \mathcal N_i$ with cavity tensors $p_{\text{g}\to\n_a}$ and $p_{h\to\n_a}$ on the boundary.

If each neighborhood $\mathcal{N}_i$ has an environment that conforms to the decomposition assumption (which means that the neighborhoods of every boundary tensor are disconnected from each other without passing through the tensors in $\mathcal{N}_i$), we can contract the cavity networks 
 layer by layer from the boundary of the entire network to $\mathcal{N}_i$, then the result is exactly $P_i$. Otherwise, when the cavity networks share connections with others, there will be no boundary of the entire network, and if we consider the dependency relationship between cavity tensors as a kind of message flow transmitted among tensors, the flow under this network structure will form a closed loop, and we need to use an iterating algorithm to determine the cavity tensors.
Since $P_i$ only depends on the ratio of the two elements of the result vector (its holds when the related tensor is multiplied by an arbitrary constant), as long as we have the relationship among the cavity tensors, all the information needed to obtain $P_i$ can be fully determined. As a consequence, we can establish a set of iterative equations of the cavity tensors and obtain their approximate values by iterating these equations.
In the statistical mechanics problems and graphical models we considered here, the tensor network contains a large number of copy tensors, so we can make use of the properties of the copy tensors and simplify the representation of cavity tensors. In more detail, consider that a tensor $\mathcal A_{i_1,i_2,\cdots, i_m}$ contracted with a copy tensor $\mathcal{I}^{i_1,\cdots, i_m}_{j_1,\cdot,j_n}$ and obtain a tensor $\mathcal{B}_{j_1,\cdot,j_n}$. We can show that the resulted tensor can be rewritten as a contraction of a vector and a copy tensor with order $n+1$:

\begin{equation}
    \begin{split}
    \label{eq:copytensor}
    \mathcal{B}_{j_1,\cdot,j_n}
    =&\sum_{i_1,i_2,\cdots i_m } \mathcal A_{i_1,\cdots, i_m} \mathcal{I}^{i_1,\cdots, i_m}_{j_1,\cdot,j_n} \\
    =& \sum_{i_1,i_2,\cdots i_m  } \mathcal A_{i_1,\cdots, i_m} (\prod_{k=1,\cdots, m} \delta_1^{i_k}\prod_{l=1,\cdots, n} \delta^1_{j_l} + \prod_{k=1,\cdots, m} \delta_2^{i_k}\prod_{l=1,\cdots, n} \delta^2_{j_l}) \\
    =& \mathcal A_{1,\cdots,1}\prod_{l=1,\cdots, n} \delta^1_{j_l} + \mathcal A_{2,\cdots,2}\prod_{l=1,\cdots, n} \delta^2_{j_l} \\
    =& \sum_{q=1,2} ( \mathcal A_{1,\cdots,1}  \delta_q^1 \delta_1^q \prod_{l=1,\cdots, n} \delta^1_{j_l} + \mathcal A_{2,\cdots,2}  \delta_q^2 \delta_2^q \prod_{l=1,\cdots, n} \delta^2_{j_l} )\\
    =& \sum_{q=1,2} (\mathcal A_{1,\cdots,1} \delta^1_q + \mathcal A_{2,\cdots,2} \delta^2_q) (\delta_1^q \prod_{l=1,\cdots, n} \delta^1_{j_l} + \delta_2^q \prod_{l=1,\cdots, n} \delta^2_{j_l})\\
    =& \sum_{q=1,2} [ (\sum_{i_1,i_2,\cdots i_m } \mathcal A_{i_1,\cdots, i_m} \mathcal{I}^{i_1,\cdots, i_m}_q )\mathcal{I}^{q}_{j_1,\cdot,j_n} ] \\
    =&  (\mathcal A_{i_1,\cdots, i_m} \times \mathcal{I}^{i_1,\cdots, i_m}_q ) \times \mathcal{I}^{q}_{j_1,\cdot,j_n},
    \end{split}
\end{equation}
where the function $\delta^a_b = 1$ when $a=b$ and 0 otherwise. The process is illustrated in Fig. \ref{fig:Cavity}(b), where we see that the cavity tensor $p_{a \rightarrow \mathcal{N}_i}$ can be equivalently represented as the contraction of a message vector $m_{a \rightarrow \mathcal{N}_i}$ and a copy tensor. This heavily reduces the space complexity of the message-passing process as we can use $2$ message vectors rather than cavity tensors during the message-passing iteration.
We notice that his simplification is valid only in the case of existing many copy or diagonal tensors (as in the statistical mechanics problems and graphical models). For the application of our method in other problems such as computing expectations in quantum systems this simplification does not hold and we need to use cavity tensors during iterations in general.

\subsection{Constructing the neighborhoods}
If tensors on the boundary of the neighborhood of node $i$, $\mathcal{N}_i$, have weaker correlations via the tensors outside $\mathcal {N}_i$, the tensor-product assumption of the environment tensor that we have made in computing marginal of $i$ would be more accurate. In this work, we characterize the correlation using the the path (e.g. $ l\to x\to y \to\cdots\to z \to k$) outside $\mathcal N_i$ (i.e. $x,y,\cdots, z\not\in\mathcal N_i$) connecting the tensors  $l$ and $k$ on the boundary of $\mathcal N_i$ to characterize the correlation between them. A simple observation would be that the longer path, the weaker correlation. If $N_i$ contains all the tensors of the tensor network, we consider the length of the path to be $\infty$, and consistently, the computation of the marginal of $i$ using the contraction of tensors in $\mathcal N_i$ is exact.
We may consider more refined measurements of correlations by considering the coupling, such as $\exp\beta(|J_{lx}|+|J_{xy}|+\cdots+|J_{zk}|)$. 

To put it simpler and more general, in this work we only consider the length of the path and define $d_{ab}(\partial \mathcal{N}_i)$ as the length of the shortest path in the corresponding graph $G$ connecting the cavity tensors $a$ and $b$ only via the tensors outside $\mathcal{N}_i$ to quantify the correlations between $a$ and $b$ without the connections inside $\mathcal{N}_i$. To determine the neighborhoods satisfying the weak connection condition, we propose to generate $\mathcal{N}_i$ of each vertex $i$ by progressively including neighboring edges of $G$ till $\partial \mathcal{N}_i$ satisfying $\mathop{\min}_{a,b \in \partial \mathcal{N}_i} d_{ab}(\partial \mathcal{N}_i)$ not smaller than a given value $R$, then we construct the tensor network corresponding to the edge derived subgraph from the included edges, which is written as $\mathcal{N}_i(R)$. 

\begin{figure}
\centering
\includegraphics[width=0.98\columnwidth,trim=0 1 0 1,clip]{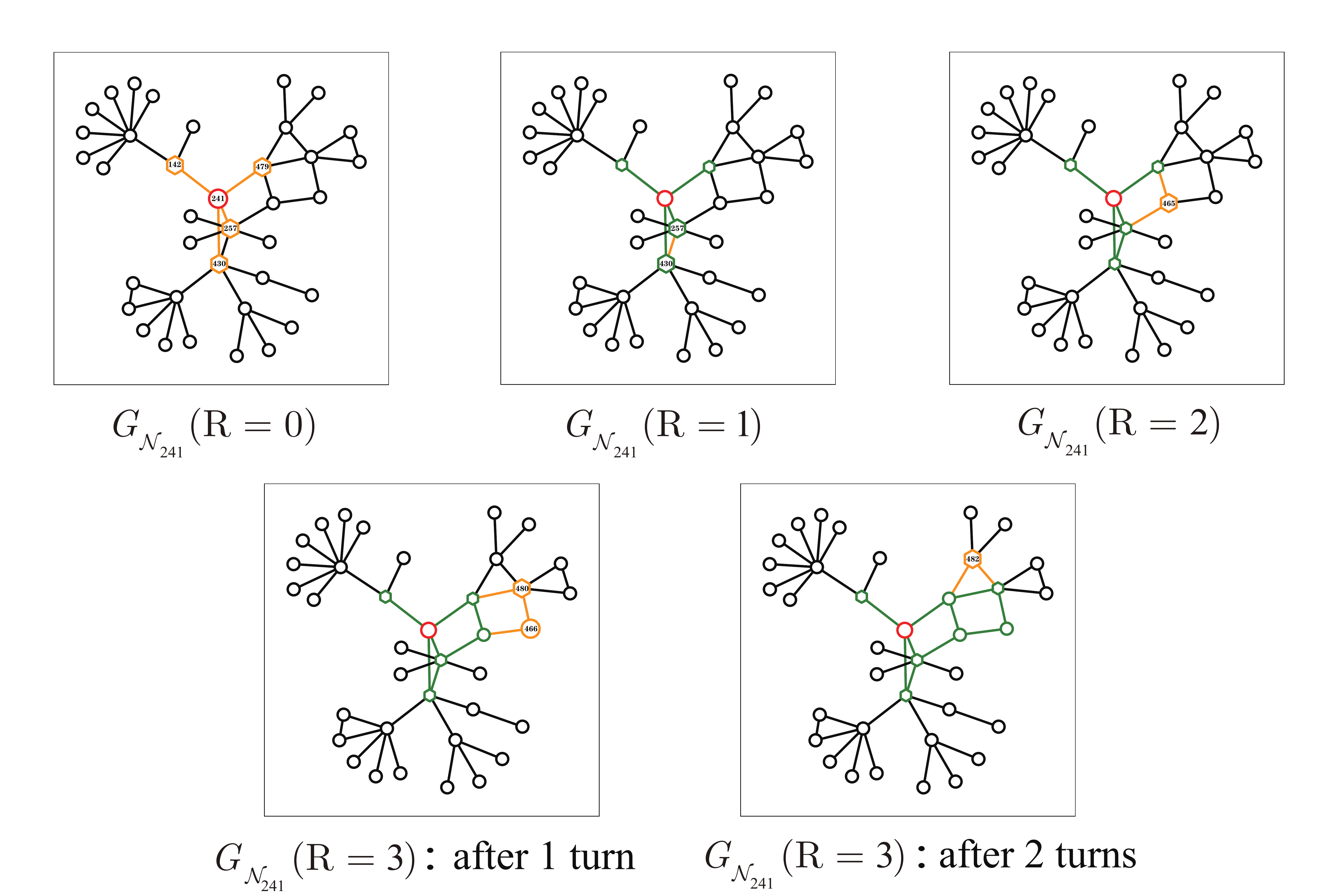}
\caption{\textbf{An example of generating the neighborhood $\mathcal{N}_i(R)$ of node 241  (the red circle) in the network of bus power system~\cite{10.1145/2049662.2049663} for $R=1,2,3$.}
Only the local subgraphs centered around node 241 are shown for clarity. The colored nodes and edges constitute $G_{\mathcal{N}_i}(R)$, where the orange elements represent newly added nodes and edges to $G_{\mathcal{N}_i}(R-1)$ or to the previous turn. Among the nodes, the dots indicate the nodes inside $G_{\mathcal{N}_i}(R)$, whereas the hexagons represent the nodes on the boundary. $R$ is the minimum distance $\min_{(ab)}d_{ab}(\partial \n_i)$ between all pairs of tensors on the boundary of the neighborhood. 
When $R=0$, $G_{\mathcal{N}_{241}}$ only comprises $241$'s direct neighbors $\left\{142,430,257,479\right\}$ and the edges connecting them and $241$.
For $R=1$, $G_{\mathcal{N}_{241}}$ needs to include edges(paths with a length of 1) between every two boundary nodes, necessitating the addition of the edge (257,430) between the only pair of connected boundary hexagons of $G_{\mathcal{N}_{241}}(R=0)$.
When $R=2$, we search for all paths with length no longer than $2$ between every boundary pair of $G_{\mathcal{N}_{241}}(R=1)$ and obtain only one such path $P = 257 \to 465 \to 479$ of length 2 between the boundary nodes 257 and 479, so $G_{\mathcal{N}_{241}}(R=2) = G_{\mathcal{N}_{241}}(R=1) \cup P$.
The procedure repeats similarly for $G_{\mathcal{N}_{241}}(R=3)$, where we add the path $P_1 = 479 \to 480 \to 466 \to 465$ in the first turn, and $P_2 = 479 \to 482 \to 480$ in the second turn. After the two turns, we obtain a neighborhood that satisfies the condition $\mathop{\min}_{a,b \in \partial \mathcal{N}_i} d_{ab}(\partial \mathcal{N}_i) \geq 3$. }
\label{fig:Neighborhood_generating}
\end{figure}

More specifically, to construct the corresponding sub-graph of $\mathcal{N}_i(R)$ (written as $G_{\mathcal{N}_i}(R)$) from a given graph $G$, we first generate $G_{\mathcal{N}_i}(R=0)$, which is simply the direct neighbors of $i$ and the edges including $i$. Then we generate $G_{\mathcal{N}_i}(R=1)$ by adding the edges shared by any two direct neighbors of $i$. On this basis, upon $G_{\mathcal{N}_i}(R=r-1)$, we generate $G_{\mathcal{N}_i}(R=r)$ by iterating the following procedure:
\begin{itemize}
    \item Find the boundary of the neighborhood, which refers to the vertices connecting nodes outside the neighborhood.
    
    \item For all the pairs of vertices on the boundary, find all the paths between them in $G \backslash G_{\mathcal{N}_i}$ with the length smaller than $R$, then add the vertices and edges on the path into the neighborhood. 
    
    \end{itemize}
We call each iteration of the above procedure a \textbf{turn}. This procedure is iterated until we obtain  $\mathcal{N}_i(R)$ from $G_{\mathcal{N}_i}(R)$ with a boundary $\partial \mathcal{N}_i$ satisfying $\mathop{\min}_{a,b \in \partial \mathcal{N}_i} d_{ab}(\partial \mathcal{N}_i) \geq R$. For example, with $G_{\mathcal{N}_i}(R=0)$ and $G_{\mathcal{N}_i}(R=1)$ being constructed, we can generate $G_{\mathcal{N}_i}(R=2)$ and $G_{\mathcal{N}_i}(R=3)$ and so on sequentially, as illustrated in Fig.~\ref{fig:Neighborhood_generating}, 
where we give an example of generating $\mathcal{N}_{241}$ in the network of bus power system~\cite{10.1145/2049662.2049663} discussed above with $R=3$.

\begin{figure}
    \centering
\includegraphics[width=0.6\columnwidth,trim=0 1 0 1,clip]{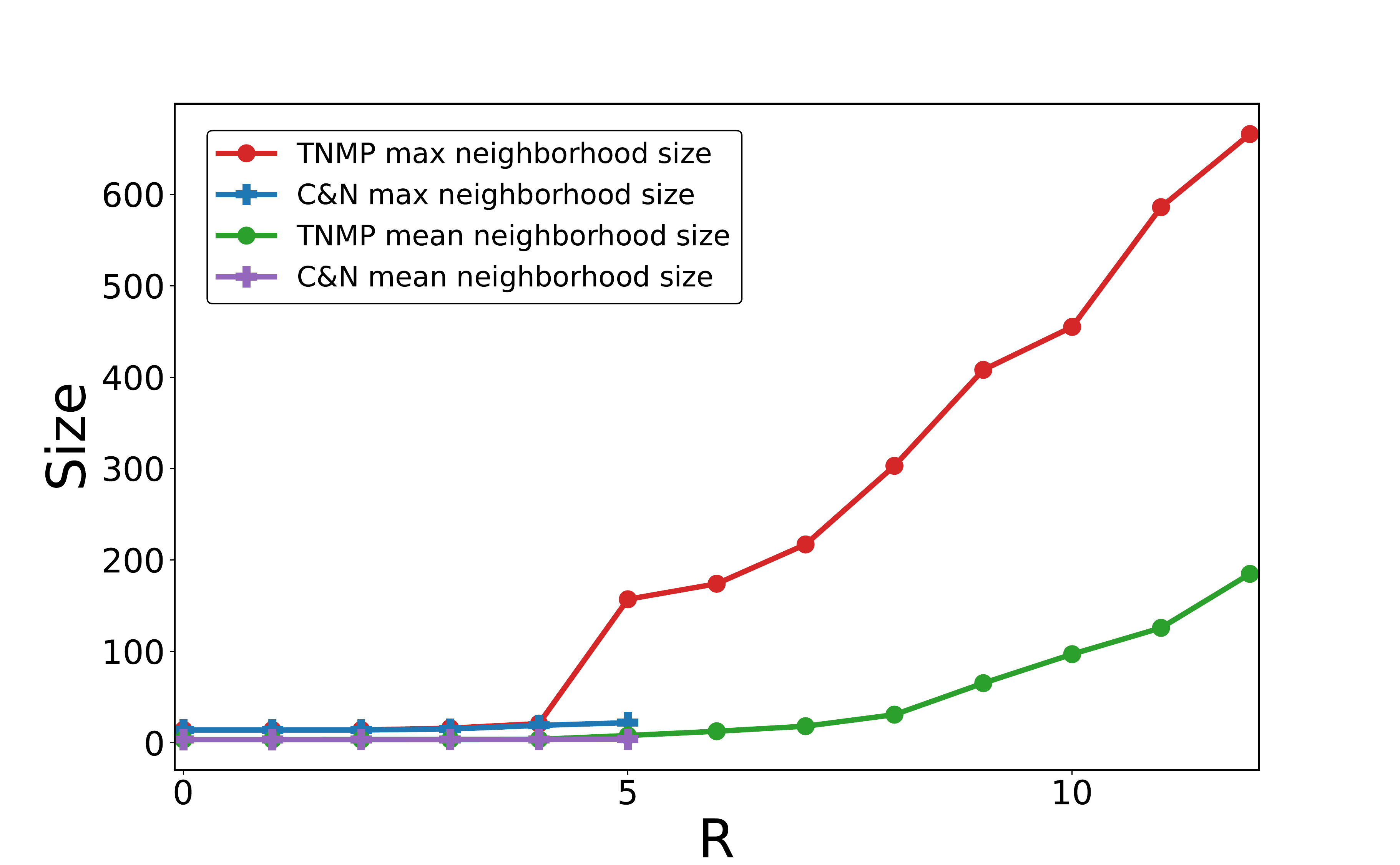}
\includegraphics[width=0.6\columnwidth,trim=0 1 0 1,clip]{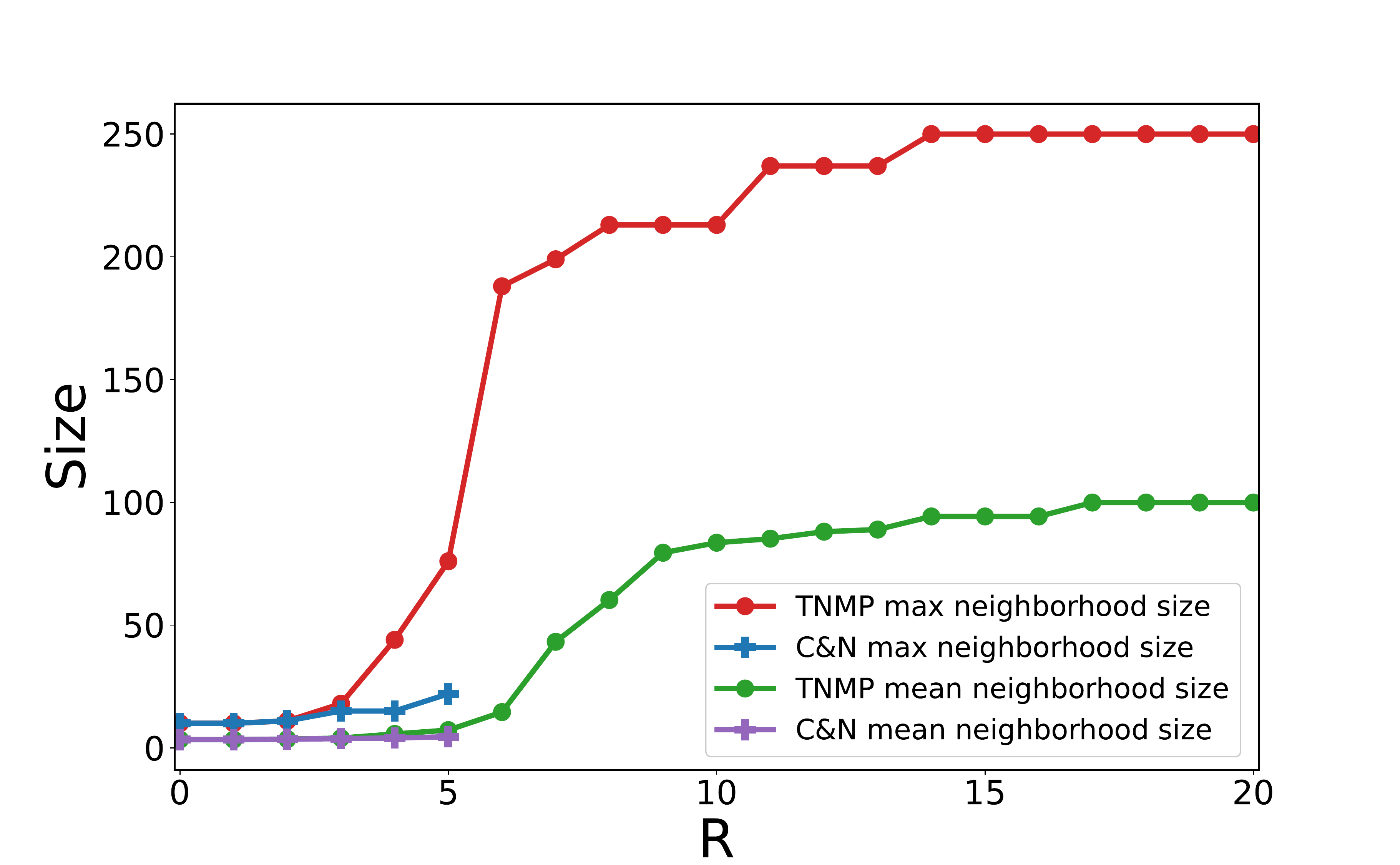}
\includegraphics[width=0.6\columnwidth,trim=0 1 0 1,clip]{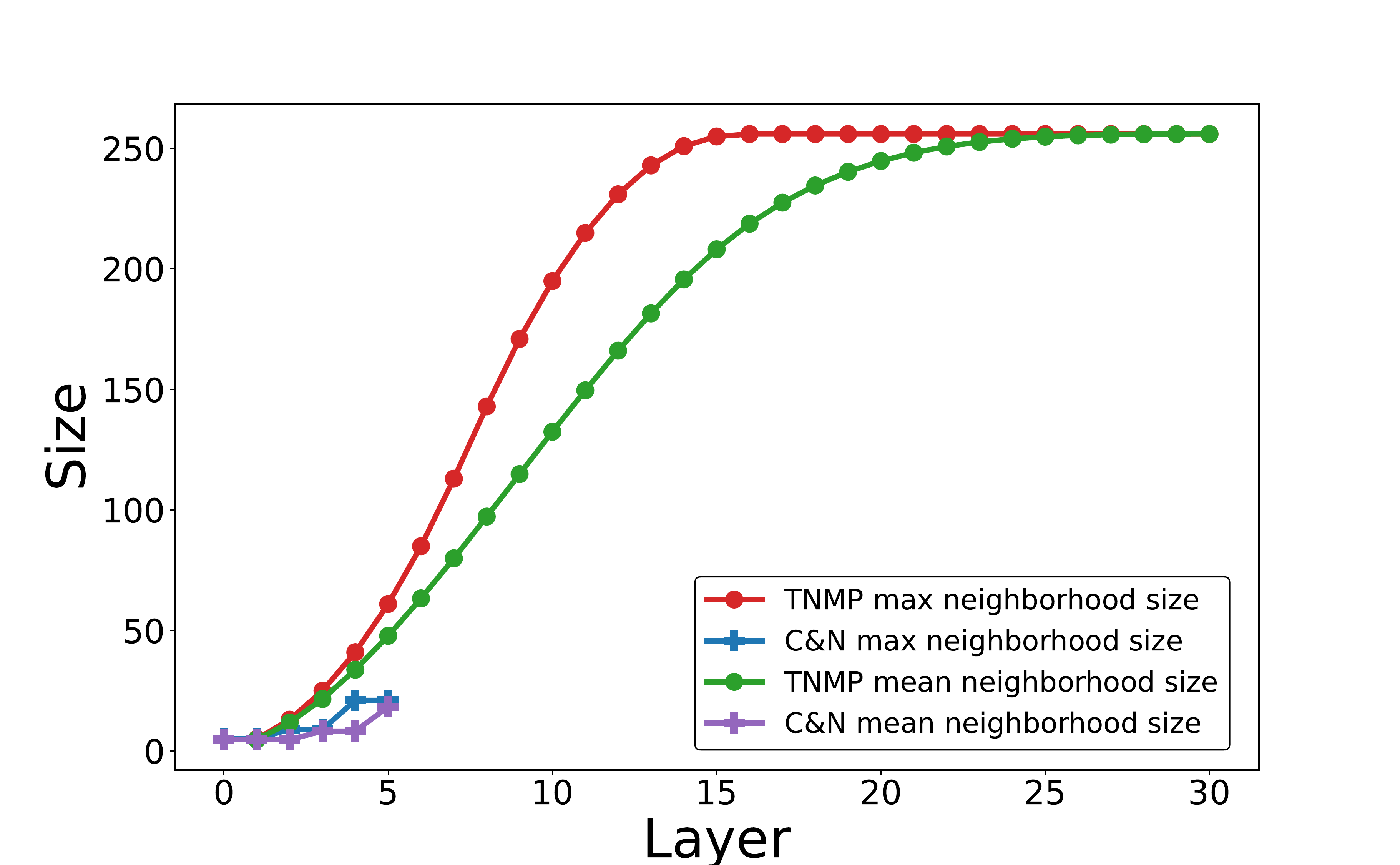}

\caption{The maximum and average neighborhood size over all variables in the graph given by our TNMP method and Cantwell and Newman's method (C\&N)~\cite{cantwell2019message}, for (Top) the synthetic graph generated using the model of~\cite{williamson2020random}; (Middle) the real-world network of bus power system~\cite{10.1145/2049662.2049663}, and (Bottom)  a 16 $\times$ 16 square lattice. In our method, $R$ is the minimum distance $\min_{(ab)}d_{ab}(\partial \n_i)$ between all pairs of tensors on the boundary of the neighborhood.
In Cantwell and Newman's method,  $R$ is the maximum length of the path under consideration between the neighbors of a node. 
}
\label{fig:size}
\end{figure}

In Fig.~\ref{fig:size} we compare the maximum neighborhood size and mean neighborhood size given by our method (TNMP) and the method of Cantwell and Newman (C\&N) on the Synthetic network composed of long and short loops, the real network of bus power system~\cite{10.1145/2049662.2049663}, and the 2D lattice. For Cantwell and Newman's method, we only report values with neighborhood sizes no larger than $20$. From the figure, we can see that the neighborhood sizes of TNMP are larger than that of Cantwell and Newman with the same $R$. This reflects the different implications of $R$, as well as different definitions of the neighborhood in our method and that of Cantwell and Newman~\cite{cantwell2019message}.

\begin{figure}
\centering
\includegraphics[width=0.9\columnwidth,trim=0 1 0 1,clip]{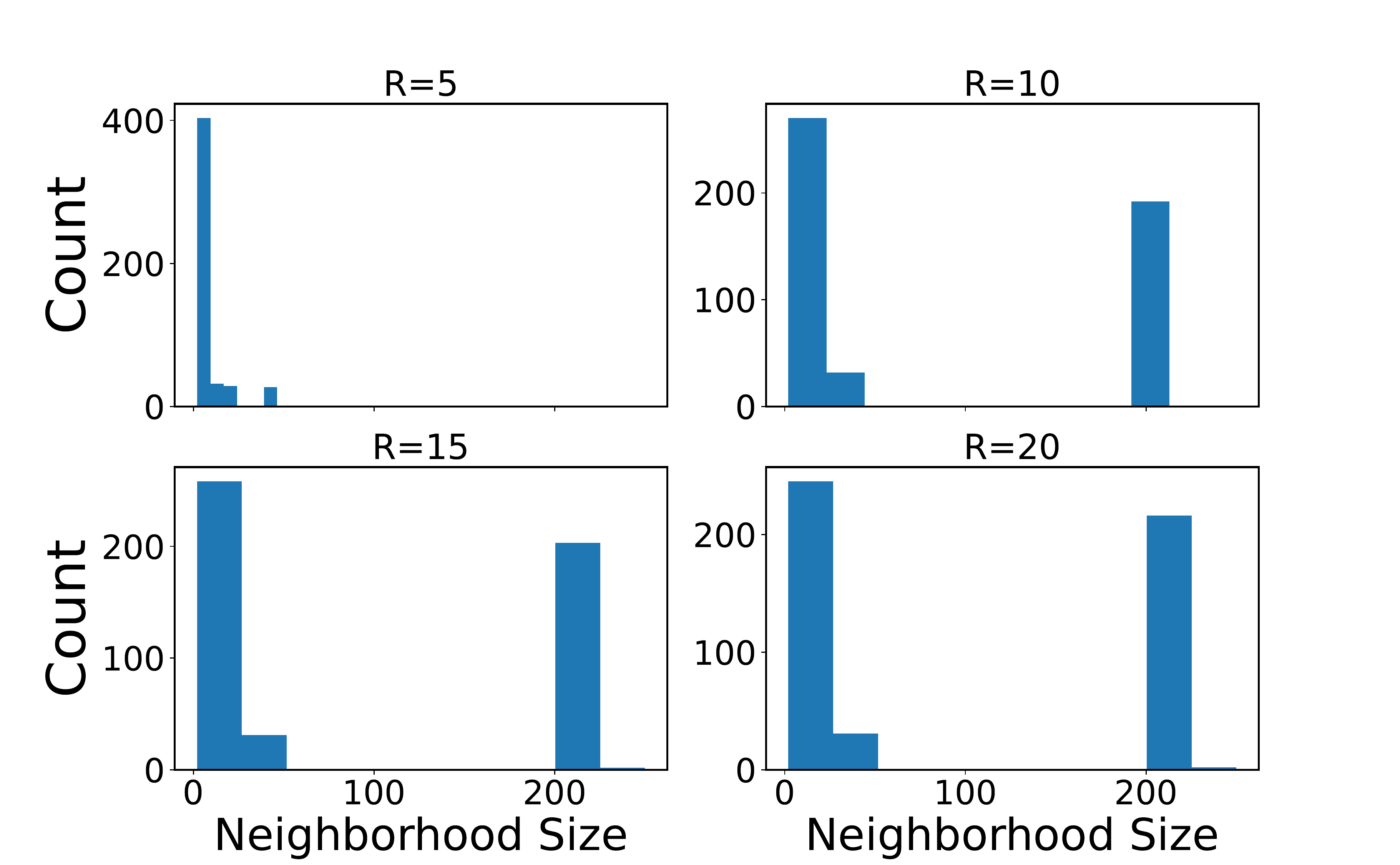}
\centering
\caption{The histogram expressing the distribution of the neighborhood size of TNMP at different R on the real-world network of bus power system~\cite{10.1145/2049662.2049663}. The $Count$ in y-axis denotes the number of nodes having the given neighborhood size.}
\label{fig:neighborhoodsize_distribution}
\end{figure}

To further display the distribution of the neighborhood sizes on different vertices of $G$, we show the histogram of the frequency of each neighborhood size given by TNMP at different R on the real-world network of bus power system~\cite{10.1145/2049662.2049663} in Fig.~\ref{fig:neighborhoodsize_distribution}.
We can see from the histogram that when $R$ is large, there are more and more large neighbors with sizes greater than $200$, which calls for an efficient algorithm for contracting the neighborhood tensor network.

\subsection{Detailed description of the TNMP process}
The pseudo-code of the TNMP algorithm on marginal calculation is listed in Tab.\ref{alg:TNMP}, where $G=(\mathcal{V,E})$ represents the graph , $J$ represents the coupling matrix, $h$ represents the external field vector, $\beta=\frac{1}{T}$ is the inverse temperature, and $\epsilon$ controls the convergence accuracy.

\begin{algorithm}[!h]
    \caption{TNMP marginal calculation}
    \label{alg:TNMP}
    \renewcommand{\algorithmicrequire}{\textbf{Input:}}
    \renewcommand{\algorithmicensure}{\textbf{Output:}}
    \begin{algorithmic}[1]
        \REQUIRE Graph $G$ with a set of vertices $\mathcal V$, coupling matrix $J$, external fields $h$, inverse temperature $\beta$, convergence criterion $\epsilon$  
        \ENSURE Marginal probability $P_i(s_i)$ of each vertex $i\in \mathcal V$    
        \STATE  Convert each node to a copy tensor, convert each field to a vector, and convert each coupling to the Boltzmann matrix to form a tensor network $T$.
        \STATE Determine the neighborhood $\mathcal{N}_i$ for each $i$.
        \STATE Generate the cavity sub-network $\mathcal{C}_{a \to i}=\mathcal{N}_a \backslash \mathcal{N}_i$ for each boundary tensor of $\mathcal{N}_i$, which is the sub-network corresponding to the edge-induced subgraph of $\mathcal{E}(G_{\mathcal{C}_{a \to i}})=\mathcal{E}(G_{\mathcal{N}_{a }}) \backslash \mathcal{E}(G_{\mathcal{N}_{i }})$.
        \STATE Initialize all the message vectors $m_{a \rightarrow i}(t=0)$ for every boundary-center pair to $(0.5,0.5)$, the number of turns of message iteration $t$ is initialized to 0.
        \WHILE{$\mathrm{diffference}=\max_{}(|m_{a \rightarrow i}(t+1)-m_{a \rightarrow i}(t)|)>\epsilon$}
            \FOR{each $i \in \mathcal{V}$}
                \FOR{each boundary tensor $a$ of $\mathcal{N}_i$}
                    \STATE Contract $\mathcal{N}_a \backslash \mathcal{N}_i$ together with the environment tensors on the boundary of $\mathcal N_a\backslash \mathcal N_i$ to get $m_{a \rightarrow i}(t+1)$.
                \ENDFOR
            \ENDFOR 
        \STATE $t=t+1$
        \ENDWHILE
        \FOR{each $i \in \mathcal{V}$}
            \STATE Contract $\mathcal{N}_i$ with converged environment tensors on the boundary of $\mathcal N_i$ to get $P_i(s_i)$.
        \ENDFOR
        \RETURN $\{P_i(s_i)\}$
    \end{algorithmic}
\end{algorithm}

\begin{figure}
\centering
\includegraphics[width=0.85\columnwidth,trim=0 1 0 1,clip]{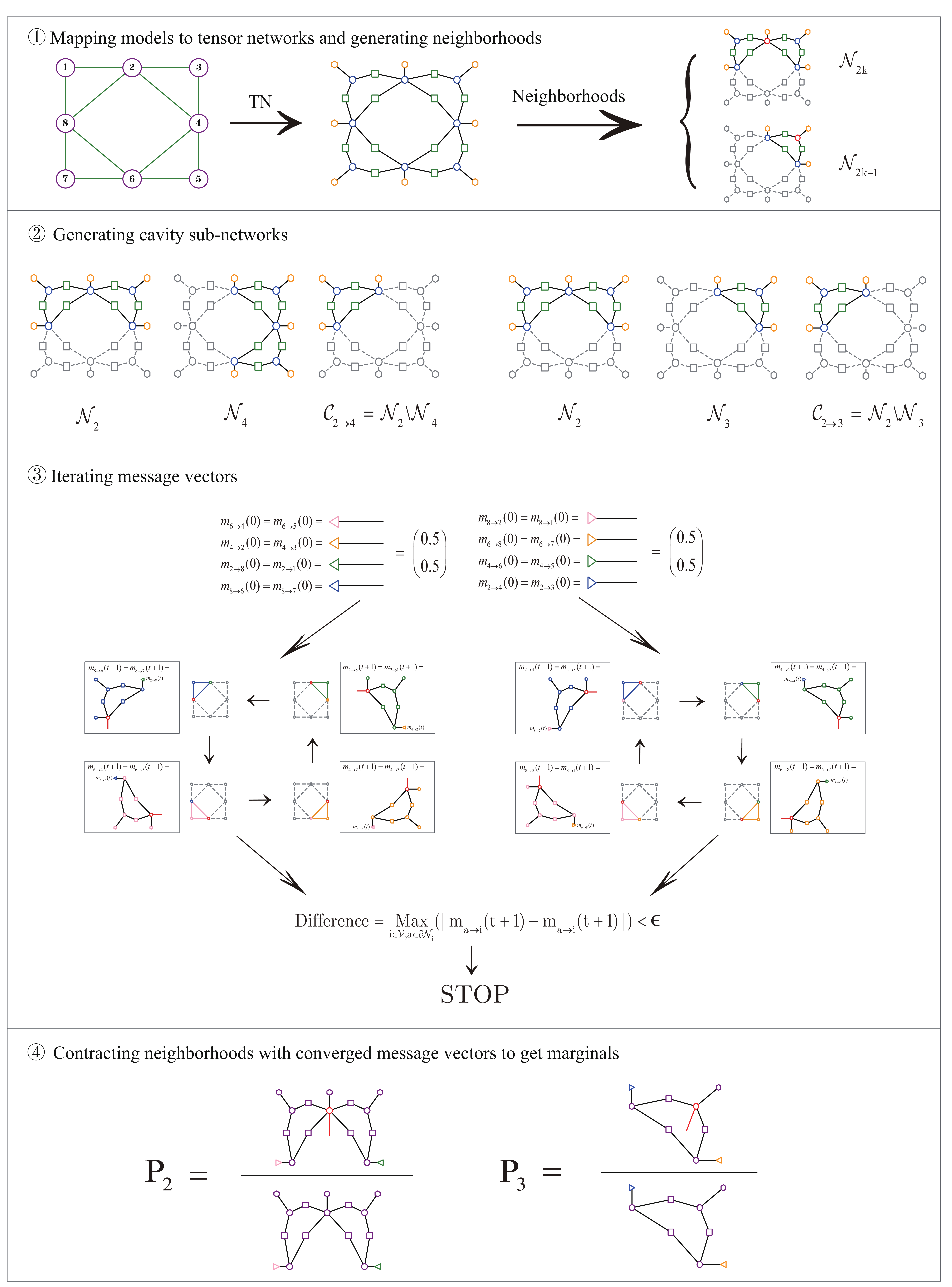}
\caption{\textbf{Detailed process of TNMP marginal calculation on a toy model with short loops.}
In the pictorial representations of tensor networks, circles represent copy tensors, hexagons represent field vectors, squares represent Boltzmann matrices, and triangles represent message vectors. (The colors of the nodes may vary across figures of different steps, but they always represent the same tensor.)
In step \ding{172}, we map the original graph onto the corresponding tensor network and generate $\mathcal{N}_i(R=1)$ for each vertex. Due to graph symmetry, there exist only two neighborhood structures: $\mathcal{N}_{2k}$ and $\mathcal{N}_{2k-1}$ for $k \in \left\{1,2,3,4\right\}$, which are shown as colored parts in the representations.
In step \ding{173}, we construct the cavity sub-network $\mathcal{C}_{i \to j}$ by selecting Boltzmann matrices that only appear in $\mathcal{N}_i$ but not in $\mathcal{N}_j$. These matrices, along with the copy tensors connected to them and the corresponding field vectors,  constitute $\mathcal{C}_{i \to j}$. There are also two structures of $\mathcal{C}$, the examples of which are also shown in colored parts.
In step \ding{174}, we iterate the message vectors through local tensor network contraction until convergence. The respective subgraph $G_{\mathcal{C}_{i \to j}}$ are displayed as colored parts on the cycle of the flowchart, alongside the corresponding tensor networks proportional to $m_{i \to j}$, where the triangles denote the message vectors from the last iteration.
In step \ding{175}, similar to Fig. \ref{fig:GTT}, $P_{i}$ is the quotient of the approximation of $Z_{i}$ and the approximation of $Z$. The triangles in the tensor network representation indicate the converged message vectors from the last iteration turn.
}
\label{fig:Process}
\end{figure}

We also give a simple pictorial illustration of the whole TNMP procedure for calculating the marginal probabilities of each variable with the connectivity denoted by the graph $G$  composed of four triangles connected end to end in a loop as shown in Fig. \ref{fig:Process} \ding{172}.

\subsection{Observable evaluations}
\begin{figure}
\centering
\includegraphics[width=0.98\columnwidth,trim=0 1 0 1,clip]{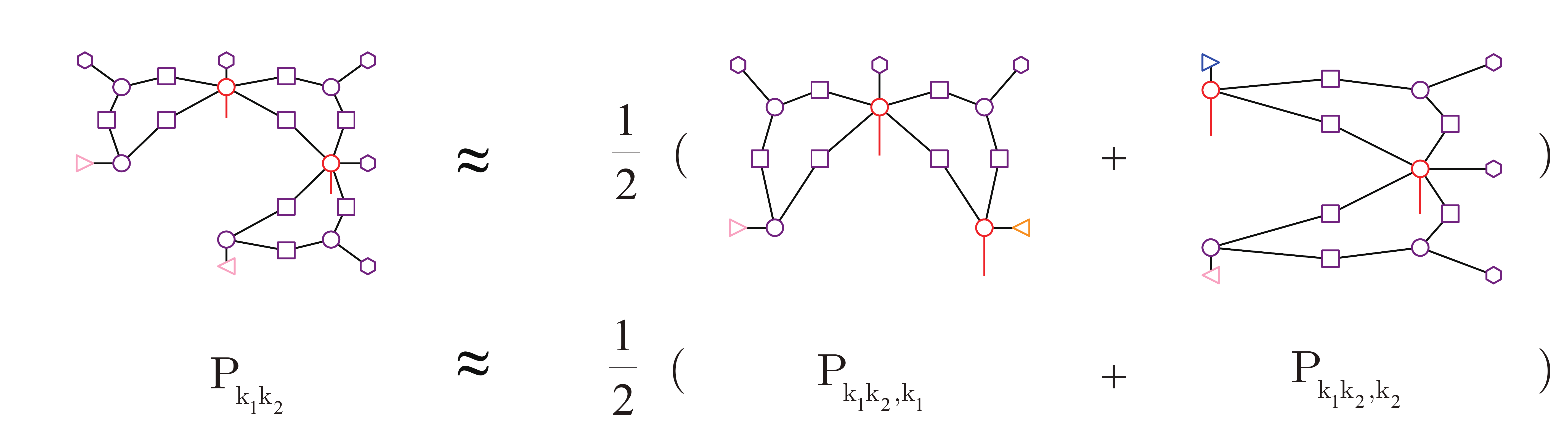}

\caption{Pictorial illustration of the approximate calculation of $P_{k_1,k_2}$ for the model in Fig. \ref{fig:Process}. The correspondence of elements in the diagram follows the representation of Fig \ref{fig:Process} \ding{175}.
}
\label{fig:U_app}
\end{figure}

With the converged cavity tensors computed at hand, the physical quantities that depend on the local contraction of the tensors network can be computed straightforwardly. For example, the magnetization $M$ is computed as 
\begin{equation}
    M = \frac{1}{|\mathcal{V}|}\sum_{i \in \mathcal{V}} [P_i(s_i=+1)-P_i(s_i=-1)],
\end{equation}
and, the internal energy $U$ can be computed as
\begin{equation}
    \begin{split}
    U &= \sum_{\s}P(\s)E(\s)\\
    &= \sum_{\s}P(\s) [\sum_{(i,j) \in \mathcal{E}} -J_{ij}s_is_j-\sum_{i \in \mathcal{V}}h_is_i]\\
    &= \sum_{(i,j) \in \mathcal{E}}[\sum_{\s}P(\s)(-J_{ij}s_is_j)] + \sum_{i \in \mathcal{V}}[\sum_{\mathbf{s}}P(\s)(-h_{i}s_i)]\\
   &= \sum_{(i,j) \in \mathcal{E}}[\sum_{s_i,s_j}P_{ij}(s_i,s_j)(-J_{ij}s_is_j)] + \sum_{i \in \mathcal{V}}[\sum_{s_i}P_i(s_i)(-h_{i}s_i)],
    \end{split}
    \label{eq:U}
\end{equation}
where the both the two-point marginal probabilities $P_{ij}(s_i,s_j)$ and single point marginals $P_i(s_i)$ can be computed efficiently using a local neighborhood tensor network contraction with the cavity tensors on the boundary.
In detail, to get $P_{ij}$, we first determine a neighborhood $\mathcal{N}_{ij}$ with two centers $i$ and $j$, and contract $\mathcal{N}_i$ as shown in Fig. \ref{fig:GTT}(d), to get $P_{ij,i}$ and repeat the same contraction for $\mathcal{N}_j$ to get $P_{ij,j}$. Then we compute the internal energy $U$ by approximating $P_{ij}$ to $\frac{1}{2}(P_{ij,i}+P_{ij,j})$ and substituting all the $P_{ij}$ into Eq. \ref{eq:U}. We also give a simple example of the calculation of $P_{k_1,k_2}$ for the model in Fig. \ref{fig:Process} in Fig. \ref{fig:U_app}.
Another possible way to compute the two-point marginal $P_{ij}(s_i,s_j)$ would be contracting the neighborhood tensor network $\mathcal N_{ij}$ containing both $i$ and $j$, and the cavity tensors on the boundary of $\mathcal N_{ij}$.

\subsection{Order and slicing for tensor network contractions}
\subsubsection{Contraction order and slcing}
To perform the contraction on a tensor network composed of a large number of tensors, we maintain a sequence of tensor pairs (e.g. $(1,2),(2,3),(3,4)
,\cdots,(n,n-1)$), and at each step of the contraction, a pair of tensors is popped from the sequence and contracted into a single tensor. This sequence of tensor pairs is termed as \textit{contraction order $\pi$}, where all the initial and intermediate tensors appear only once. The computational cost of the entire contraction process heavily depends on the contraction order. The minimum cost of the contraction among all possible contractions is determined by the exponential of the \textit{tree width} of the network \cite{Markov2008}, which is defined as the maximum \textit{bag size} under the optimal t{tree decomposition}.

The computational cost of the contraction process with order $\pi$ is usually characterized using the \textit{space complexity} (\textbf{sc}($\pi$)) and the \textit{time complexity} (\textbf{tc}($\pi$)). The space complexity is the maximum sum of dimensions of all snapshots during the contraction specified by the order $\pi$, and is usually estimated by the dimension of the maximum intermediate tensor during the whole contraction. The time complexity is the logarithm of the summation of the number of scalar multiplications in each tensor-pair contraction, which is equal to the product of dimensions of indices involved in each contraction step of $\pi$. As demonstrated in \cite{Markov2008}, $\min_{\pi} \textbf{sc}(\pi)$ of a tensor network is equal to the treewidth of its corresponding graph. So with the optimal contraction order, the computational cost of TNMP depends on the treewidth of the corresponding graph.

However, it is an NP-hard problem to determine the treewidth of the network and to find the contraction order that gives the lowest complexity. Heuristics algorithms have been developed to find a contraction order with low computational complexity, for example, greedy strategies that choose a pair of tensors that produces the largest ratio between the dimension of the input tensors and the dimension of the output tensor of the contraction. In ~\cite{kourtis_fast_2019} an algorithm is proposed to first find a balanced partitioning of the tensor network with a small cut size, then find contraction orders to the sub-tensor networks corresponding to the partitions of the graph individually. A series of works following the balanced-partitioning-based strategy has been applied to many problems such as solving constraint satisfaction problems~\cite{kourtis_fast_2019} and quantum computer simulations~\cite{gray_hyper-optimized_2021,pan_simulation_2022}. Recently, it has been shown that the contraction order obtained by searching the contraction tree using e.g. simulated annealing gives an even better quality~\cite{kalachev_multitensor_2022}. In this work, we chose an open-source implementation of a heuristic algorithm \textit{opt-einsum} \cite{daniel2018opt}. The algorithm finds a contraction order by performing a recursive, depth-first search, together with branching, greedy selection, as well as a balanced-partitioning heuristic. We note that other algorithms and non-public packages for finding the contraction order give similar performance, e.g. using~\cite{pan_simulation_2022}.

In addition to contracting the neighborhoods in TNMP, in the numerical experiments, to obtain exact results to measure the error of different algorithms by contracting the whole tensor network, the space complexity (i.e. exponential of the tree-width) of the underlying tensor network could exceed the storage limit of the physical device, so we use the \textit{dynamic slicing} method to achieve exact contraction of excessively large tensor networks. Since what tensor network contraction does is to multiply the elements of each tensor under given values of indices and then sum over these products at all the possible values of indices, in the slicing method, some indices are chosen to explicitly sum over ``manually'' rather than being included as tensor dimensions, which turns a single contraction into many independent, easier sub-contractions. With the selection of the sliced indices, the slicing method fixes the values of the sliced indices and enumerates the combinatorial values of corresponding indices to generate a new tensor network, on which a sub-contraction is performed. Then the summation of the results of these $2^{|\textbf{sliced indices}|}$ sub-contractions is equal to the result of total tensor network contraction.
In the exact marginal calculation, we directly call an open-source implementation Cotengra \cite{gray_hyper-optimized_2021} to find the sliced indices and the contraction order after slicing, which is based on the picture of \textit{contraction tree} and perform the search by simultaneously interleaving subtree reconfiguration with slicing.

\subsubsection{Approximate contraction, and the trade-off between the contraction error and message passing error}
In our algorithm TNMP, when the neighborhood and the corresponding local tensor network for computing marginals and cavity tensors are very large (e.g. with a large $R$),  the space complexity of contraction would exceed the upper limit of physical devices, and the time  complexity of the local contraction would also be significant, we need to consider approximate contractions.
The challenge of approximate contraction of the neighborhood tensor network is that the connectivity of the network is irregular, so the traditional methods of tensor renormalizations and boundary matrix product states do not apply. Fortunately, in recent years various methods of arbitrary contractions have been proposed~\cite{pan_contracting_2020,gray_hyper-optimized_2022}. In this work we use the CATN method which was proposed by some of us in \cite{pan_contracting_2020}. In the CATN method, the tensors are represented, compressed, and contracted using the matrix product states (MPS) in the canonical form, analogous to the density matrix renormalization method. With the MPS representation, we can always control the sizes of the intermediate tensors under the limit of the physical devices by reducing their virtual bond dimension $\chi$ and physical bond dimension $D$.

In CATN, every tensor is represented in a \textit{standard MPS} satisfying: (1) its bond dimensions $D$ and $\chi$ is not larger than the given parameter $\hat{D}$ and $\hat{\chi}$, (2) in the canonical form, (3) its indicators are rearranged to make the index to be contracted at the tail or head of the index sequence and (4) shares no more than one common edge with any one of the other tensors (also represented as MPSes) in the network. 
Any tensor can be approximated as such a standard MPS by QR decompositions and singular value decompositions~\cite{schollwock2011density,orus2014practical}. More specifically, for the four requirements of standard MPS, (1) can be achieved by SVD truncation, (2) by QR decompositions, (3) and (4) by ``swap" and  ``merge" operations, whose principle is to restore a part of the MPS to a raw tensor and redo the SVD truncation.

We call the above approximate conversion from an initial raw tensor or an intermediate general MPS to a standard MPS as \textit{Preparation}, and like the exact tensor network contraction, a \textit{Contraction} step means the operation of summing over the common index of two tensors. Then what CATN does is repeat (1) Preparation and (2) Contraction according to the contraction order until there is only one result tensor in the network.
In CATN, the larger value of bond dimensions, the more faithfully MPS can represent the original tensor, bringing a more accurate approximation and smaller errors. 

If we increase the size of the local tensor network to contract, it is more difficult to accomplish the local tensor network contraction, leading to a larger difference between the CATN method with given computational resources (e.g. with a fixed $\hat{D}$ and $\hat{\chi}$) and the exact result, producing \textit{contraction error}. However, at the same time, the larger local tensor network usually has a larger $R$ value hence reducing the correlation between the tensors on the boundary of the local tensor network (by increasing the length of the loops connecting the tensors via tensors outside the local tensor network), hence reducing the \textit{message-passing error}. Obviously, the contraction error is an increasing function of the neighborhood size while the message passing error is a decreasing function of the neighborhood size. So, the size of the neighborhood sub-graph plays a tradeoff between the contraction error and the message passing error. As a result, the change of the total error with the neighborhood size is neither a monotonically decreasing function nor a monotonically increasing function. Instead, it depends on the computational resources the CATN algorithm uses.
Moreover, both the bond dimension and the neighborhood size are positively correlated with computational complexity, where the bond dimension measures the computational cost of achieving more accurate local CATN and the neighborhood size measures the cost of achieving more accurate global message passing. So under a constant limitation of given calculation conditions, we can achieve smaller errors by adjusting the three parameters $\hat{D}$, $\hat{\chi}$, and $R$.

To illustrate the trade-off between the contraction error from CATN and the message passing error, we investigate the performance of our method on the spin glass model on $16\times 16$ square lattice with random interactions and random fields. In the square lattice, no matter how we define the neighborhood for node $i$, the minimum distance between pairs on the boundary is always very small. In this sense, the quality of the rank-one environment is always inaccurate, so the performance of the results purely depends on the size of the neighborhood, as a monotonically increasing function. In our evaluation, since the minimum distance does not increase with the size of the neighborhood, we determine the neighborhood of node $i$ by including its neighbors and the neighbors of neighbors, etc, i.e. layer by layer.

\begin{figure}
\centering
\includegraphics[width=0.7\columnwidth,trim=0 0 0 10,clip]{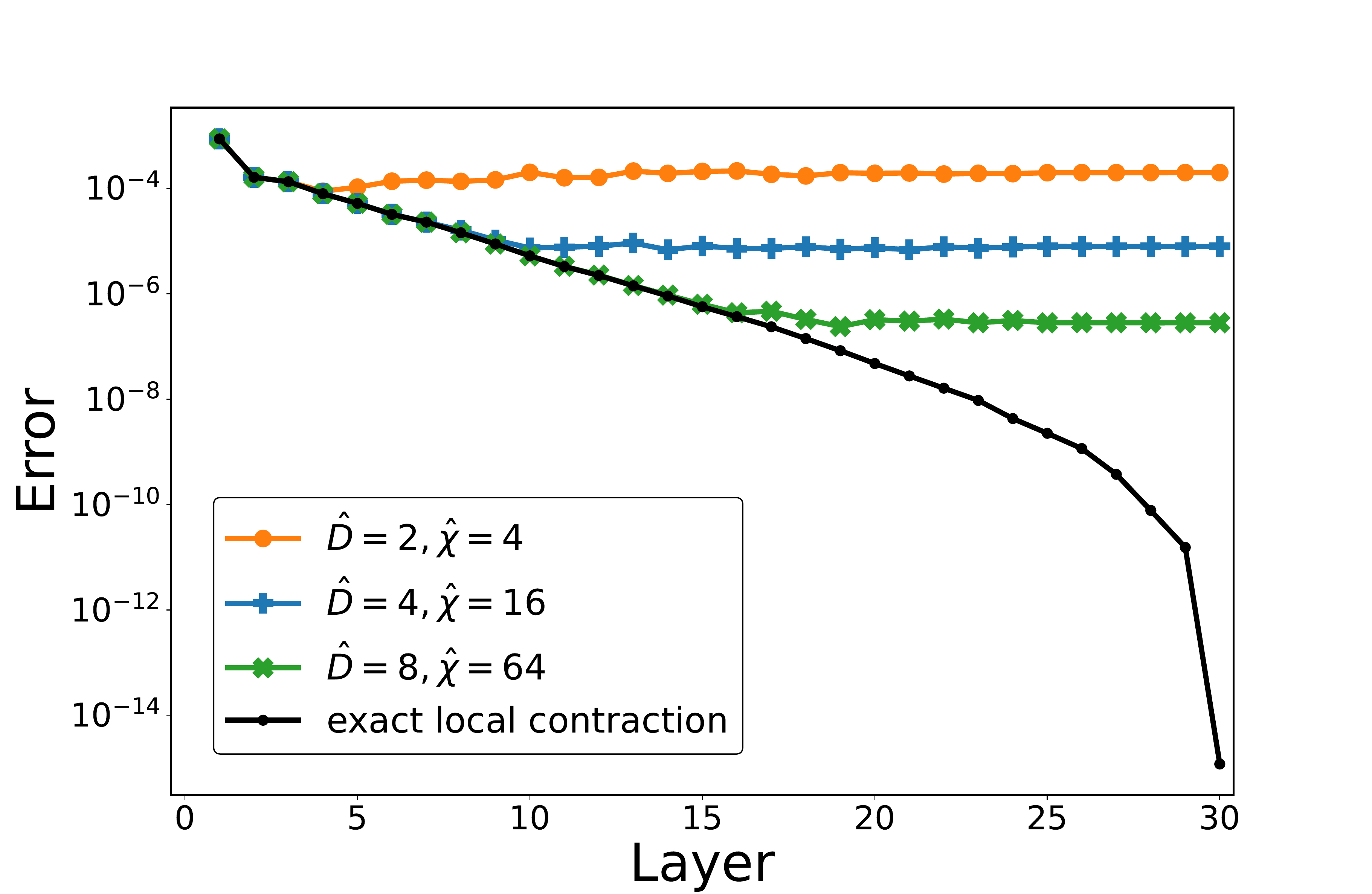}
\caption{Error $\frac{1}{n}\sqrt{\sum_{i}{(M_i - M^{exact}_i)^2}}$ of TNMP in computing mean marginal as a function of the number of $layers$ ($G_{\mathcal{N}_i}$ is defined as the vertex induced subgraph of $\mathcal{V} = \left\{v \in \mathcal{V}(G) | d(v,i) \leq R \right\}$) with local neighborhood contracted by CATN with different bond dimensions $\hat D$ and $\hat \chi$, on a random-interaction and random-field Ising model on 16 $\times$ 16 square lattices at the fixed temperature $T=2.0$. The coupling $J \sim  Bernoulli(0.5) \times 2 - 1$ and the external field $h \sim  N(0,\frac{\pi}{200})$.}
\label{fig:2d}
\end{figure}

\begin{figure}
\centering
\includegraphics[width=0.9\columnwidth,trim=0 1 0 1,clip]{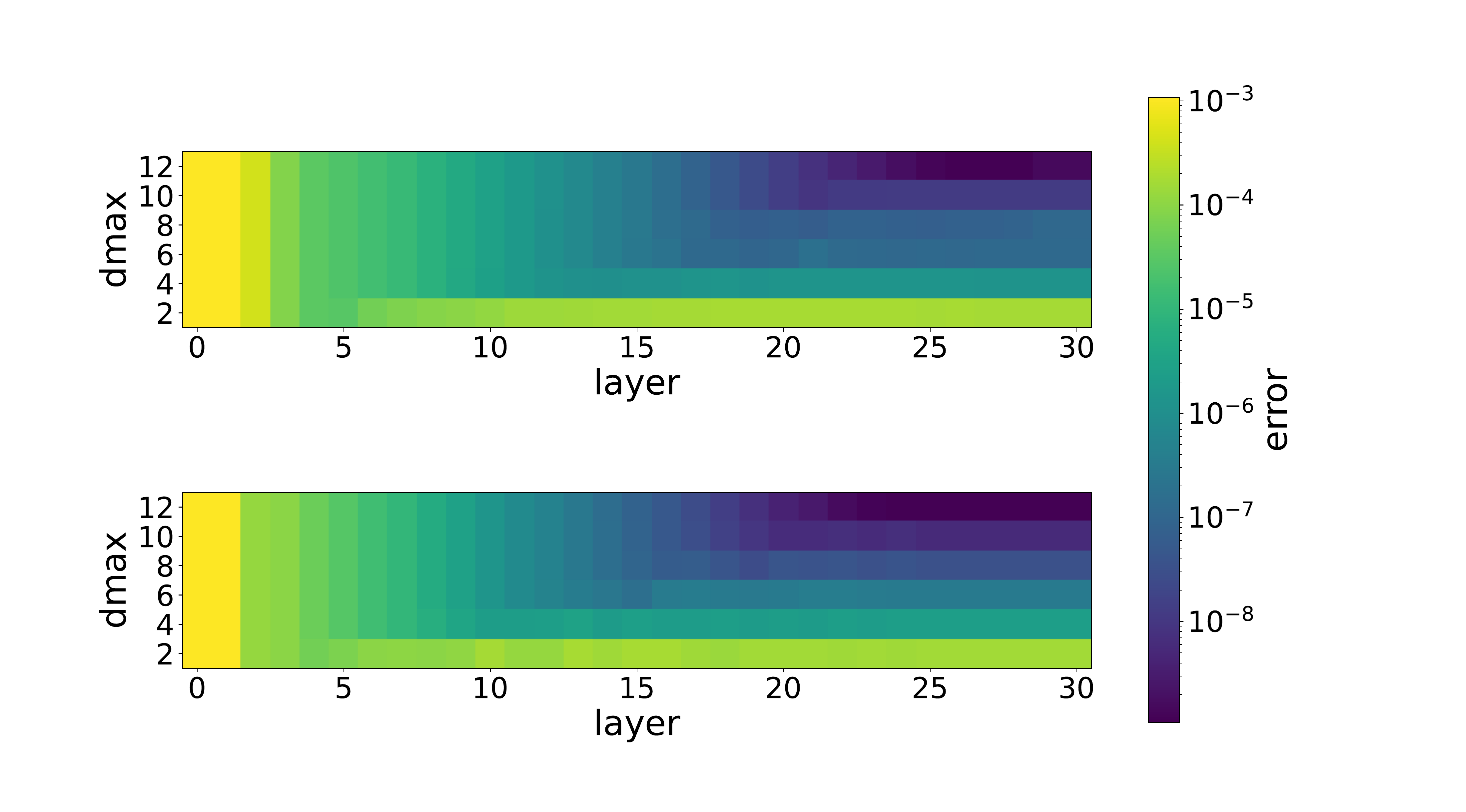}
\centering
\caption{{Heat map of the error of magnetization $E = \frac{1}{n}\sqrt{\sum_{i}{(M_i - M^{exact}_i)^2}}$ vs $layer$ and the maximum bond dimension $dmax$ of CATN} on 16 $\times$ 16 square lattice at the fixed temperature $T=2.0$. (Top): The result on the ferromagnetic Ising model with a constant external field, where all couplings take $+1$ and all external fields $h_i= 0.1$.
(Bottom): The result on the random-interaction and random-field spin glass model, where each coupling takes $+1$ and $-1$ with an equal probability, and each external field $h_i$ follows $h_i \sim  N(0,\frac{\pi}{200})$.}
\label{fig:2d16_heat}
\end{figure}

As shown in Fig.~\ref{fig:2d}, with the number of layers increasing, in the beginning, the total error is almost equal to the message passing error which can be estimated from the total error with an exact local contraction. Then as the message passing error decreases to a level comparable to the CATN error, the local contraction error begins to dominate the total error, so the curve starts to move away from the curve corresponding to the exact local contraction and finally reaches a plateau with a slow variation of the total error with the number of layers. We term the total error reached when the CATN contraction curve separates from the exact contraction curve the \textit{critical error}.
For small bond dimensions of CATN, the final plateau is higher than the critical error, so by making a trade-off between errors, TNMP reaches a lower error than both pure message passing and total tensor network approximate contraction.
For large bond dimensions, the CATN error is always small and very close to the critical error, but at the right end of the plateau, TNMP can reach the same level of error of total tensor network approximate contraction because the neighborhood almost spans the whole tensor network. The heat map of the magnetization errors is shown in Fig.~\ref{fig:2d16_heat} for both a ferromagnetic model and a spin glass model.

\subsection{Convergence of TNMP}
To make the result of the message passing algorithm \textit{consistent}, it is necessary for the message passing algorithm to converge - otherwise, different runs of the algorithm give different results. However, in general, there is no guarantee for a message-passing algorithm to converge. While they usually converge at high temperatures, some of them do not converge at low temperatures. To analyze the convergence properties of TNMP, we can abstract the local contraction with old messages to obtain new ones as update equations among messages and consider the iteration of messages as the evolution of a linear system expanded at a trivial fixed point. Then the convergence of the iteration process can be analyzed using linear stability analysis of the trivial fixed point. In this view, the linearized message passing equations correspond to the (generalized) \textit{nonbacktracking operator}~\cite{krzakala2013spectral,zhang2017spectral,Kirkley_2021}, linearization of the update equations deviating from a trivial fixed point. Using the theory of linear system stability, we can obtain the convergence characteristics of TNMP on a given model from the eigenvalues of the generalized nonbacktracking matrix. But in this work we will not investigate the corresponding generalized nonbacktarcking matrix in detail to study quantitatively the convergence and fixed points of TNMP.

\begin{figure}
\centering
\includegraphics[width=0.7\columnwidth,trim=0 1 0 1,clip]{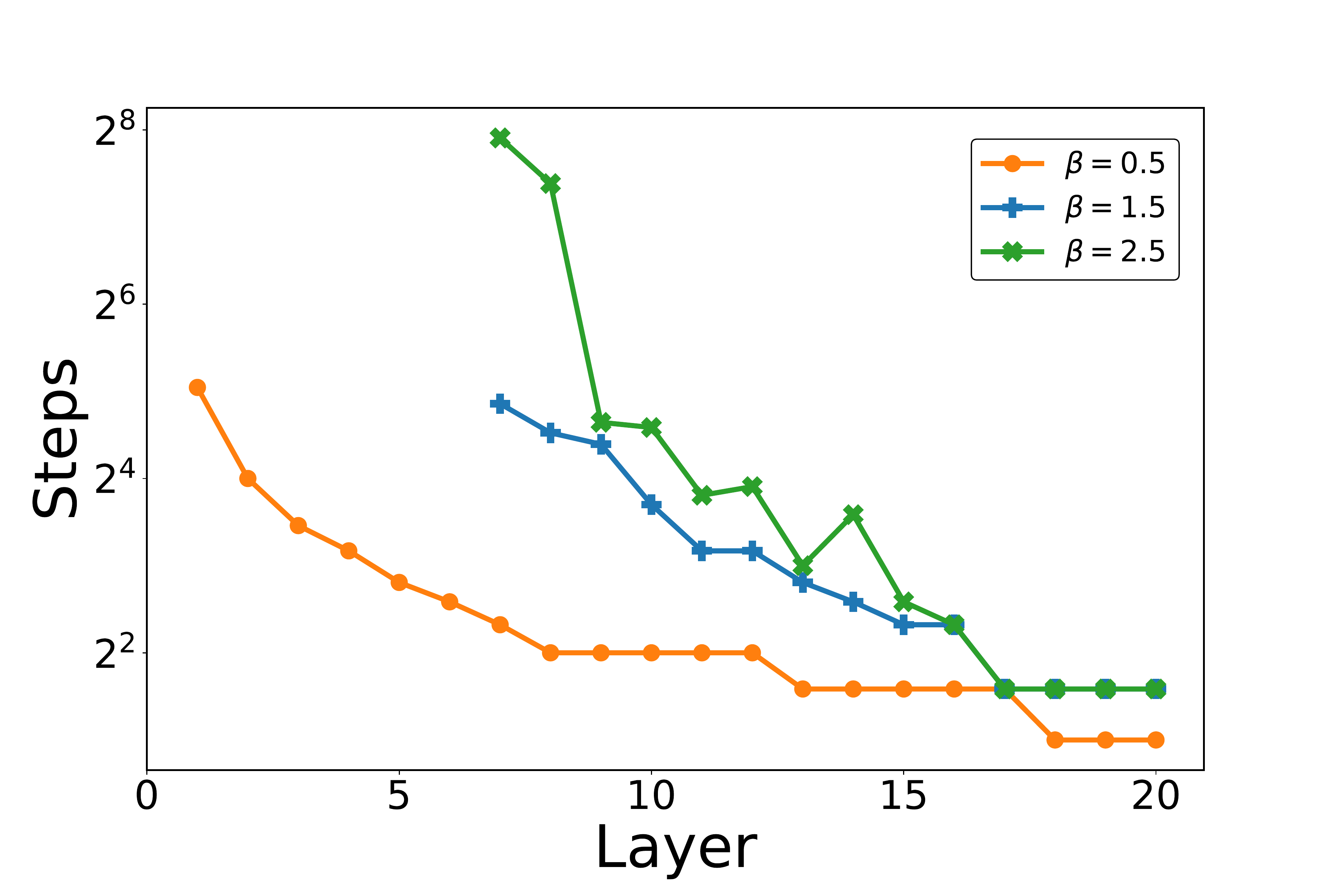}
\includegraphics[width=0.7\columnwidth,trim=0 0 0 75,clip]{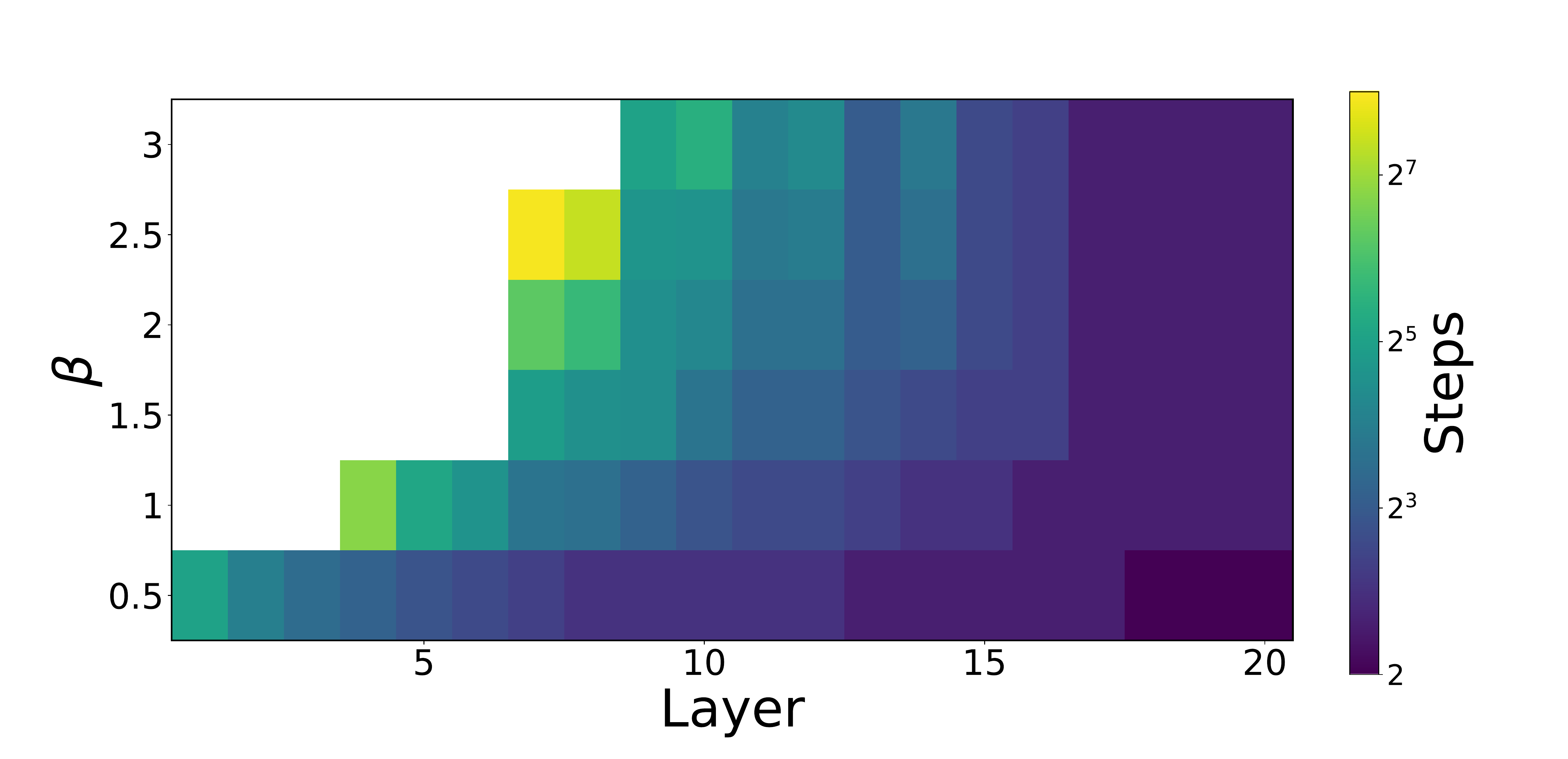}
\centering
\caption{\textbf{The number of iterations (steps) before TNMP converges} on the random-interaction and random-field Ising model behavior on 16 $\times$ 16 square lattice at different temperature $T=1/\beta$, the convergence criterion $\epsilon = 10^{-6}$. Each coupling takes $+1$ and $-1$ with equal probability, and each external field $h_i$ follows a Gaussian distribution with mean $0$ and variance $\pi/200$. (Top) Convergence steps vs $layer$ ($G_{\mathcal{N}_i}$ is defined as the vertex induced subgraph of $\mathcal{V} = \left\{v \in \mathcal{V}(G) | d(v,i) \leq R \right\}$) under different inverse temperature $\beta$. (Bottom) Heat map of the steps vs $layer$ and $\beta$, where the white area represents that the iteration cannot reach convergence under the corresponding parameters.}
\label{fig:2d16_step}
\end{figure}

To study the convergence property of TNMP, we use the simple argument that the larger neighborhood we use in TNMP, the smaller correlations among the cavity tensors on the boundary, resulting in better convergence properties of TNMP. We illustrate this using an example of TNMP for an Edward-Anderson Ising spin glass model on a $16\times 16$ lattice. We compare the number of iteration steps that TNMP requires to converge with a different layer in the neighborhood construction (the larger number of layers, the larger neighborhood) at different temperatures. The results are shown in Fig. \ref{fig:2d16_step}.  We can see that as the number of layers (as well as the neighborhood size) increases, TNMP converges with fewer iterations at the same temperature, and converges at lower temperatures with a fixed number of iteration steps.
In other words, the larger neighborhood, the better convergence properties of TNMP. When compared with the belief propagation and the generalized belief propagation algorithms on the 2D EA spin glass model as investigated in \cite{dominguez2011characterizing}, TNMP can converge in a broader range of $\beta$.

\subsection{More results on the performance of TNMP}
\begin{figure}
\centering
\includegraphics[width=0.7\columnwidth,trim=0 1 0 1,clip]{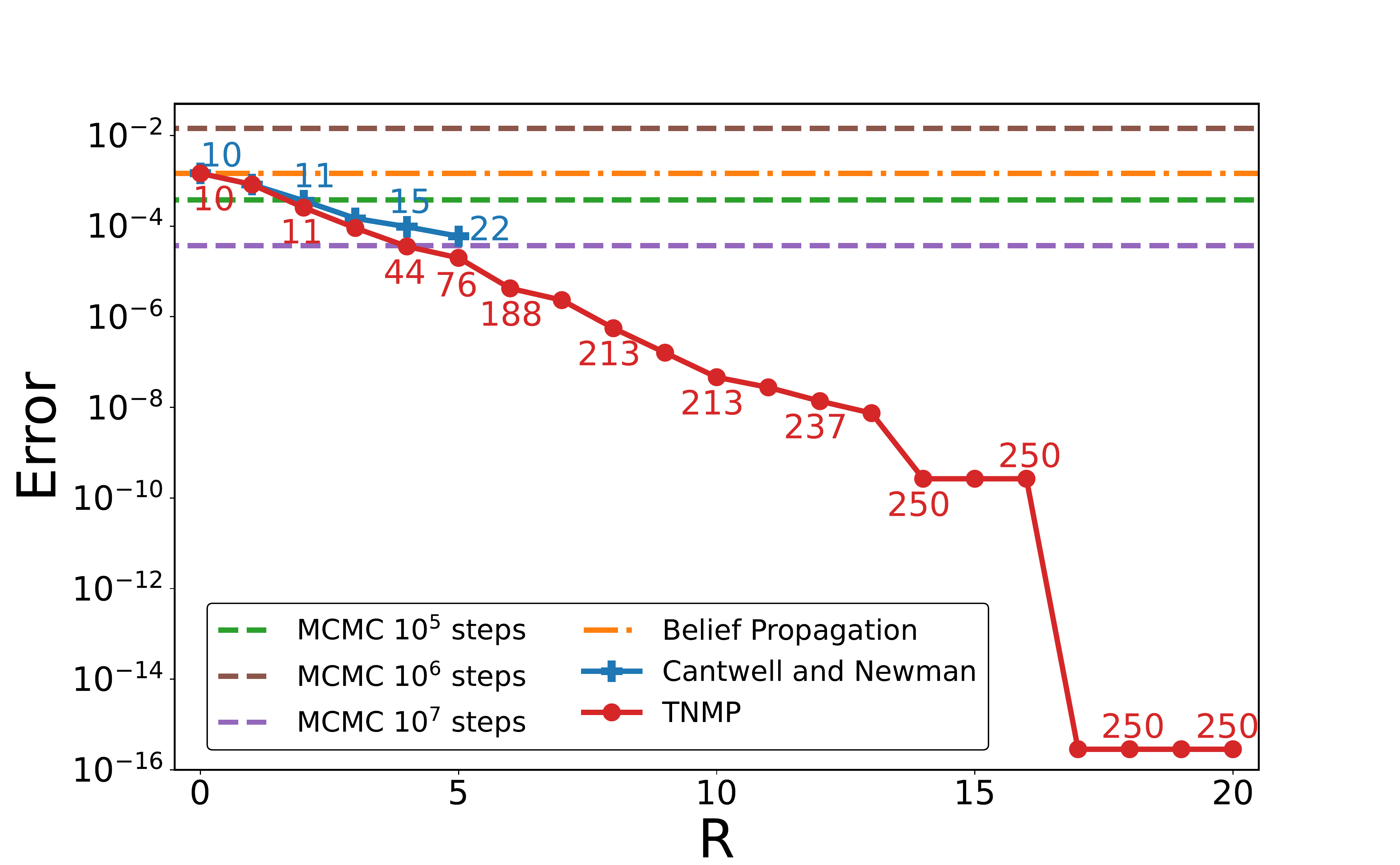}
\includegraphics[width=0.7\columnwidth,trim=0 1 0 1,clip]{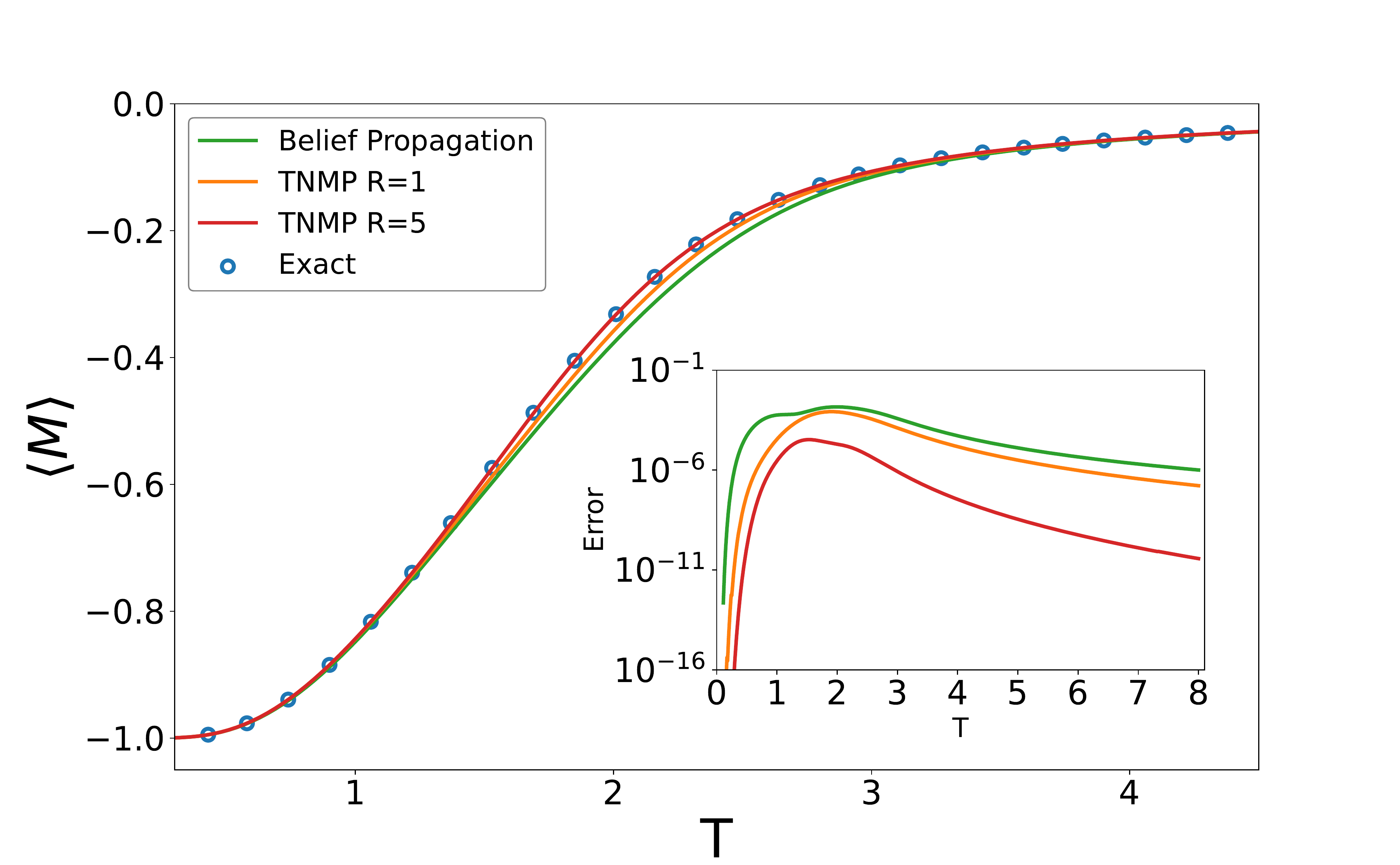}
\centering
\caption{\textbf{Behavior of TNMP for a ferromagnetic Ising model on the real-world network of bus power system~\cite{10.1145/2049662.2049663}.} The system is ferromagnetic, with couplings $J_{ij}=1$ and external fields $h_i= 0.1$. (Top) Error of magnetization $E = \frac{1}{n}\sqrt{\sum_{i}{(M_i - M^{exact}_i)^2}}$ given by various methods vs $R$ at the fixed temperature $T=2.0$. The numbers labeled in the figure indicate maximum neighbor size $\max_{i}|\n_i|$ with a given $R$. In our method, $R$ is the minimum distance $\min_{(ab)}d_{ab}(\partial \n_i)$ between all pairs of tensors on the boundary of the neighborhood. 
In Cantwell and Newman's method,  $R$ is the maximum length of the path under consideration between the neighbors of a node. 
(Bottom)The average magnetisation $M$ given by TNMP under different $R$ (which is belief propagation when $R=0$) vs temperature $T=\frac{1}{\beta}$. Insets: Error of magnetization $E$ given by TNMP with different $R$ values at different temperature .}
\label{fig:494bus_ferro}
\end{figure}

In this section, we give additional results about performance of TNMP on the real-world network of bus power system~\cite{10.1145/2049662.2049663}. In top panel of Fig.~\ref{fig:494bus_ferro}, we plot the error of magnetizations as a functio of $R$, and compare with Cantwell and Newman's method and MCMC results. The results are qualitatively similar to that the TNMP behavior on the spin glass model on this graph. The bottom panel of Fig.~\ref{fig:494bus_ferro} displays the mean magnetization obtained using TNMP compared with the belief propagation and the exact results. We can see that with $R=5$, TNMP performs coincides very well to the exact results.

\end{document}